\documentclass[intlimits,twoside,a4paper]{article}

\usepackage{graphicx}

\usepackage[cp1251]{inputenc}

\usepackage{cmpj3}

\issue{2018}{21}{2}{23601}
\doinumber{10.5488/CMP.21.23601}

\title[Solvent composition effects on the properties of ionic solutions]
{Effects of ion concentration and solvent composition on the properties of  
water-methanol solutions of NaCl.
NPT molecular dynamics computer simulation results} 

\author[M. Cruz Sanchez, J. Gujt, S. Sokolowski, O. Pizio]{M. Cruz Sanchez\refaddr{label1}, J. Gujt\refaddr{label2}, S. Sokolowski\refaddr{label3}, O. Pizio\refaddr{label4}\footnote{On sabbatical leave from Instituto de Qu\'{i}mica de la UNAM, 
corresponding author: oapizio@gmail.com.}}

\addresses{
\addr{label1} Instituto de Qu\'{i}mica, Universidad Nacional Aut\'{o}noma de M\'{e}xico,\\
Circuito Exterior, 04510, Cd. de M\'{e}xico, M\'{e}xico
\addr{label2} Department of Chemistry, University of Paderborn, Warburger Strasse 100,
33098 Paderborn, Germany
\addr{label3}Department for the Modelling of Physico-Chemical Processes, 
Maria Curie-Sklodowska University, \\ Lublin 20-614, Poland
\addr{label4}Instituto de Investigaciones en Materiales, Universidad Nacional 
Aut\'{o}noma de M\'{e}xico,\\ Circuito Exterior, 04510, Cd. de M\'{e}xico, M\'{e}xico 
}

\date{Received January 1, 2018, in final form March 29, 2018}

\begin{document}
\maketitle

\begin{abstract}
Isothermal-isobaric  molecular dynamics simulations 
are used to examine the microscopic structure and other  properties
of a model solution consisting of NaCl salt dissolved in  water-methanol mixture. 
The SPC/E water model and the united atom model for methanol are combined 
with the force field for ions by Dang [J. Amer. Chem. Soc.,  1995, \textbf{117}, 6954] 
to describe the entire system.
Our principal focus is to study the effects of two variables, namely, the
solvent composition and ion concentrations on the solution's density, on the 
structural properties, self-diffusion coefficients of the species and the 
dielectric constant. Moreover, we performed a detailed analysis
of the first coordination numbers of the species. 
Trends of the behaviour of the average number of hydrogen bonds between solvent 
molecules are evaluated.


\pacs 61.20.-p, 61.20-Gy, 61.20.Ja, 65.20.Jk

\end{abstract}

\section{Introduction}

Theoretical studies of ionic solutes in water have generated an enormous amount of literature 
during the recent decades. However, several problems still 
require a comprehensive  solution, particularly due to the lack of a complete 
understanding of water as a solvent. 
The recent progress in the area has been related to the application of
computer simulation methods to the models of various charged species in aqueous 
media with different degrees of sophistication, see, e.g.~\cite{ivo,vrabec1}. 

Less studies have been devoted to the properties of ionic solutes, even
simple salts, in combined solvents that include water and an organic component.
These systems exhibit a wide range of physicochemical phenomena. They are 
of great importance  for basic and applied research, but not easy to deal with due to their complexity. 
 The amount of experimental data in this area is much less 
comprehensive compared to ionic aqueous solutions.

For specific purposes of our project, it is worth mentioning that in a large
number of works the molecular dynamics computer
simulations have been applied to mixtures of water with various solvents. The mixtures of water with alcohols,
with DMSO, DMF, acetonitrile, acetone and glymes seem to be most frequently studied,
see, e.g.~\cite{galicia1,galicia2,vargas,dominguez1,dominguez2,perera} and references therein. 
Actually, practical needs, besides academic
interest,  determine the choice of systems for experimental research and 
for computer simulations.

The addition of ionic species to each of the above mentioned combined, 
e.g., two-component, solvents results in complex 
systems with peculiar structural, thermodynamic, dynamic and dielectric properties
dependent on the ion concentration, on the type of co-solvent, and on the solvent composition, 
besides temperature and pressure.
Various ionic solutions comprising several co-solvents have been investigated using
experimental techniques and computer simulations,
see, e.g.~\cite{patey4,tembe,ewa1,ewa2,ewa3,ewa4,ewa5,ewa6,ewa7,ewa8,ewa9,takamuku1,takamuku2,takamuku3,takamuku4,takamuku5,taha,bouazizi}.

In particular, salt-induced phase separation of aqueous mixtures of a water-miscible organic solvent has been the subject of experimental and computational 
studies~\cite{takamuku1,takamuku2,takamuku3,takamuku4,takamuku5,taha}.
On the other hand, ionic solutions of halides with alkali and doubly valent cations in 
water-methanol solvents of various composition do not exhibit this kind of phase 
separation~\cite{ewa1,ewa2,ewa3,ewa4,ewa5,ewa6,ewa7,ewa8,ewa9}. Interestingly,
the potassium carbonate seems to be the only salt capable of splitting aqueous methanol 
solutions into two liquid phases~\cite{iliuta}.

The principal objective of the present work, as a part of an ampler project,
is to investigate the microscopic structure and other properties
of sodium chloride ionic solutions at different ion concentrations with water-methanol
mixed solvent in the entire range of compositions. In spite of impressive efforts
and success of long-term studies coming from the laboratory of 
Hawlicka~\cite{ewa1,ewa2,ewa3,ewa4,ewa5,ewa6,ewa7,ewa8,ewa9}, there are several issues
that require further exploration of this type of solutions. Some of these issues
will be discussed below.

Our study is performed in the framework of molecular dynamics computer simulations.
The structure of solutions is given in terms of different descriptors,
namely, a brief account of the radial distribution functions is provided, 
and the first coordination numbers are discussed. 
In addition, we would like to analyze the conditions 
under which the cluster formation of ions occurs. With this aim, we analyze the time evolution of 
the number of clusters and other related
distributions. The structure of the maximal cluster formed at a certain ion
concentration and solvent chemical composition is visualized.  Insights into the
dynamic properties are obtained by exploring the mean square displacements as a 
function of time and the resulting self-diffusion coefficients of the species. 
Changes of hydrogen bonds statistics are analyzed. Finally, we 
discuss the behaviour of the dielectric constant of the solutions upon the changes 
in the solvent composition and ion concentration. 

At this initial stage  of the project, we performed all the analyses using a single 
combination of the force field models for water, methanol and ions as well as a single 
combination rule for the cross interactions.
Sodium chloride, well soluble in water, has been chosen here as a test case
to investigate other salts and co-solvents in our future work.
Moreover, our study is restricted to room temperature and ambient pressure  (1~bar).
Only the changes of salt concentration and the chemical composition of 
the solvent are investigated.
Wider, more precise and profound insights into the thermodynamics of this type 
of solutions, the analyses of an ampler set of dynamic and dielectric properties, require 
 additional explorations.

\section{Model and simulation details}

Prior to discussing the simulation techniques, we would like to comment on the modelling used
in the previous studies of alkali halides solutions with water-methanol solvent.
The system in question, i.e., NaCl-water-methanol, is complicated for simulations as it involves 
a model for each compound. Moreover, certain combination rules should be used for cross interactions. 

In almost all cases, Hawlicka et al.~\cite{ewa1,ewa2,ewa3,ewa4,ewa5,ewa6,ewa7,ewa8,ewa9}
used the three site Bopp-Jancso-Heinzinger (BJH) model for water~\cite{bopp}
and a three-site Palinkas-Hawlicka-Heinzinger (PHH) model for methanol~\cite{palinkas}.
The ion-solvent and ion-ion potentials were derived from ab initio calculations by fitting
the energies to analytical expressions, see, e.g.~\cite{bako}. In some
studies, the Lorentz-Berthelot (LB) combination rule was used.  
All the simulation results published by Hawlicka group have been obtained using 
either microcanonical (NVE) or canonical (NVT) ensemble, the simulation box sizes 
in each procedure have been adjusted to the experimental density values, that
unfortunately have not been specified but in a few cases turned out to be in terms of simulation box length.

Recent report by Bouazizi and Nasr~\cite{bouazizi}  
combined the flexible SPC model for water and the flexible six-site model for methanol 
(apparently, the OPLS all-atom model)
with Smith and Dang model for ions~\cite{smith}. The LB combination
rules were used. The results of the procedure follow from the
NVT simulations using Berendsen control for temperature, where a rather
small number of particles (three ion pairs and 250 solvent molecules) was simulated. 
A few runs were performed with 1024 solvent molecules.

The most recent study of the system in question performed in~\cite{hasse}
involve the SPC/E model for water~\cite{spce} as  well as the models for methanol
and for ions developed in the laboratory of Vrabec~\cite{vrabec2,vrabec3}. Again,
the LB combination rules were employed. However, a correction into the cross-diameter rule was adopted for water-methanol. The simulations
were performed in the isothermal-isobaric (NPT) ensemble. 

Our calculations have been performed in the isothermal-isobaric (NPT) ensemble 
at 1~bar, and at a temperature of 298.15~K. We used the GROMACS 
software package~\cite{gromacs}, version 5.1.2.  
For methanol, we used the united atom model~\cite{jorgensen}.
The SPC/E model was taken for water \cite{spce}.
Lennard-Jones parameters for sodium and chloride ions were chosen according 
to the force field \cite{dang}.
The Lorentz-Berthelot combination rules were used to determine cross parameters. 
The nonbonded interactions were cut off at 1.4~nm. 
The long-range electrostatic interactions were handled using the
particle mesh Ewald method implemented in the GROMACS software package (fourth 
order, Fourier spacing equal to 0.12) with a precision of $10^{-5}$.
The van der Waals tail correction terms to the energy and pressure were taken into account.
In order to maintain the geometry of the water and methanol  molecules, the LINCS 
algorithm was used.

As concerns the procedure, a periodic cubic simulation box was set up for each system. 
The GROMACS genbox tool was employed to randomly place all particles in the 
simulation box.
To remove possible overlaps of particles introduced by the procedure of
preparation
of the initial configuration, each system underwent an energy
minimization using the steepest descent algorithm implemented in the GROMACS
package. Minimization was followed by a 50~ps NPT equilibration run at 298.15~K and 1~bar using
a timestep of 0.25~fs.
We used the Berendsen thermostat and
barostat with $\tau_T = 1$~ps and $\tau_P = 1$~ps during equilibration.
Constant value of $4.5\cdot10^{-5}$~bar$^{-1}$ for the compressibility of the mixtures was
employed. In the case of pure methanol solvent, the compressibility
was taken to be $1.2\cdot10^{-4}$~bar$^{-1}$.
The V-rescale thermostat and Parrinello-Rahman
barostat with $\tau_T = 0.5$~ps and $\tau_P = 2.0$~ps
and the time of step 2~fs were used during the production runs.
Statistics for each mole solvent composition and various ions
concentration for any of the properties were collected over
several 10~ns NPT runs, each starting from the last configuration of the
preceding run. The time extension for each series of calculations will be mentioned below
in the appropriate place but not less than 70~ns.

In order to compare the  density of the solutions with experimental results given on the
molality scale, see the subsection
\ref{Density} below, the total number of particles 
(ions and molecules) in the box ranged
from $\approx 2500$ to $\approx 3700$, the number of NaCl ion pairs  was in the 
interval from 0 to 40.

On the other hand, while exploring the composition changes in the solvent, 
the total number of solvent molecules 
was kept fixed at 3000. Three series of calculations correspond  to
``inserting'' the $N_{\alpha}$ ($\alpha = \text{Na, Cl}$) of
positive and negative ions ($N_{\alpha} = 20, 40$ and 60) into the box containing 3000 solvent
molecules. Hence, the ion fraction, $X_{\text{ion}}$, [$X_{\text{ion}}=2N_{\alpha}/(N_{\text m}+N_{\text w}+2N_{\alpha})$]
is equal to $X_{\text{ion}} \approx 0.013$, $X_{\text{ion}} \approx 0.038$ and $X_{\text{ion}}\approx 0.0625$, 
respectively.
The composition of the solvent was being changed in each series of calculations; 
it is described by the mole fraction of methanol molecules $X_{\text m}$ [$X_{\text m}=N_{\text m}/(N_{\text m}+N_{\text w})$].
In the case $X_{\text m} = 0$, i.e., in a pure water solvent, the prepared ionic solutions correspond to
the concentration 0.366~mol/dm$^3$, 1.081~mol/dm$^3$ and 1.77~mol/dm$^3$, 
respectively.
Additional series of calculations have been performed for the salt-free case with 3000
molecules at different $X_{\text m}$. In this case, in the absence of ions, it was sufficient
to terminate the runs at the time of 60~ns.

\section{Results and discussion}

\subsection{Density of solutions} \label{Density}

We begin with the description of the effect of the density of solutions on the solvent composition and on the ion 
concentration. The simulation  results are compared with the experimental data for sodium chloride 
aqueous solutions and with the data reported by Takenaka and Takemura~\cite{takenaka} 
for solutions involving a solvent of two species, water and methanol. All the data are collected
in table~\ref{tab:table1} below.

\begin{table}[!t]
   \caption{A comparison of the experimental data for density of NaCl solutions in
    a water-methanol solvent of different composition with NPT MD results. In the first
    column, the experimental results are derived from~\cite{romankiw}. 
    All other columns contain experimental data by Takenaka and Takamura~\cite{takenaka}.}
    \label{tab:table1}
\vspace{2ex}\centering
        \begin{tabular}{c c c| c c c| c c c}
        \hline\hline
      $X_{\text m}=0.000$ & & & $X_{\text m}=0.129$& & & $X_{\text m}=0.307$ && \\
     molality & $\rho_{\text{exp}}$ & $\rho_{\text{MD}}$ & molality & $\rho_{\text{exp}}$ & $\rho_{\text{MD}}$ & molality & $\rho_{\text{exp}}$ & $\rho_{\text{MD}}$  \\
      mol/kg & kg/m$^3$ & kg/m$^3$ & mol/kg & kg/m$^3$& kg/m$^3$ &  mol/kg & kg/m$^3$& kg/m$^3$  \\

     0.5664 & 1020.0 & 1020.25 &  0.0000 & 963.1 & 965.59 & 0.0000 & 923.3 & 922.01 \\
     0.8953 & 1032.2 & 1031.41 &  0.2292 & 972.3 & 974.14 & 0.1149 & 927.8 & 926.26 \\
     0.9175 & 1033.1 & 1032.3  &  0.3825 & 978.4 & 979.67 & 0.1816 & 930.4 & 928.59 \\
     2.3521 & 1084.0 & 1076.26 &  0.5629 & 985.3 & 985.87 & 0.2554 & 933.2 & 931.16 \\
     4.5417 & 1153.0 & 1130.36 &  0.8496 & 996.2 & 995.66 & 0.3695 & 937.5 & 935.13 \\
\hline\hline
      $X_{\text m}=0.571$ & & & $X_{\text m}=0.917$ & & & $X_{\text m}=1.000$ &&\\

     molality& $\rho_{\text{exp}}$ & $\rho_{\text{MD}}$ & molality& $\rho_{\text{exp}}$ & $\rho_{\text{MD}}$ & molality& $\rho_{\text{exp}}$ & $\rho_{\text{MD}}$ \\
     mol/kg & kg/m$^3$& kg/m$^3$ &  mol/kg & kg/m$^3$& kg/m$^3$& mol/kg & kg/m$^3$& kg/m$^3$\\
    0.0000 & 866.6 & 865.81 &  0.0000 & 800.5 & 804.85 & 0.0000 & 786.5 & 791.30 \\
    0.0994 & 870.5 & 869.46 &  0.0794 & 804.0 & 808.02 & 0.0674 & 789.6 & 794.25 \\
    0.1474 & 872.4 & 871.19 &  0.1005 & 805.0 & 808.90 & 0.0958 & 790.9 & 795.68 \\
    0.2071 & 874.6 & 872.69 &  0.1234 & 806.0 & 809.83 & 0.1169 & 791.9 & 796.40 \\
    0.3686 & 877.9 & 876.10 &  0.1702 & 808.0 & 811.67 & 0.1426 & 793.0 & 797.54 \\
\hline\hline
        \end{tabular}
\end{table}

The principal conclusion made from the data shown in the table is that the model used in the
present study leads to the results that favourably agree with experimental data of 
\cite{takenaka} in a wide 
interval of molality at various compositions, namely, at $X_{\text m}= 0.129$; 0.307; 0.571 and 0.917.
At a high fraction of methanol as a co-solvent, the discrepancy between the simulation results
and experimental data is more pronounced.
However, it is of interest to perform similar, but more extensive, comparisons for two limiting 
cases of a solvent composition, namely, for pure water as a solvent ($X_{\text m}=0$) and
for pure methanol as a solvent ($X_{\text m}=1$), since the experimental data are available.
In the case of pure water as a solvent, we performed simulations for NaCl solutions at
a molar salt concentration reported in table~2 of \cite{laaksonen} using a much larger 
number of molecules in the box,  as mentioned in the previous section. Our simulation data are
given in the first column of the above table. The experimental values for these conditions 
(converted to molality scale) were derived by interpolation from the data presented by
Romankiw et al.~\cite{romankiw}. The reason for doing that is due to the fourth line in 
table~3 of ~\cite{laaksonen} being incorrect, the work by Anderson et al.~\cite{anderson} 
(cite Nu. 36 of \cite{laaksonen}) does not contain any data concerning the density of NaCl aqueous
solutions. In fact, our simulation results agree reasonably well with experimental data
for low and intermediate values of molality.    

\begin{figure}[!t]
\centering
\includegraphics[width=0.47\textwidth]{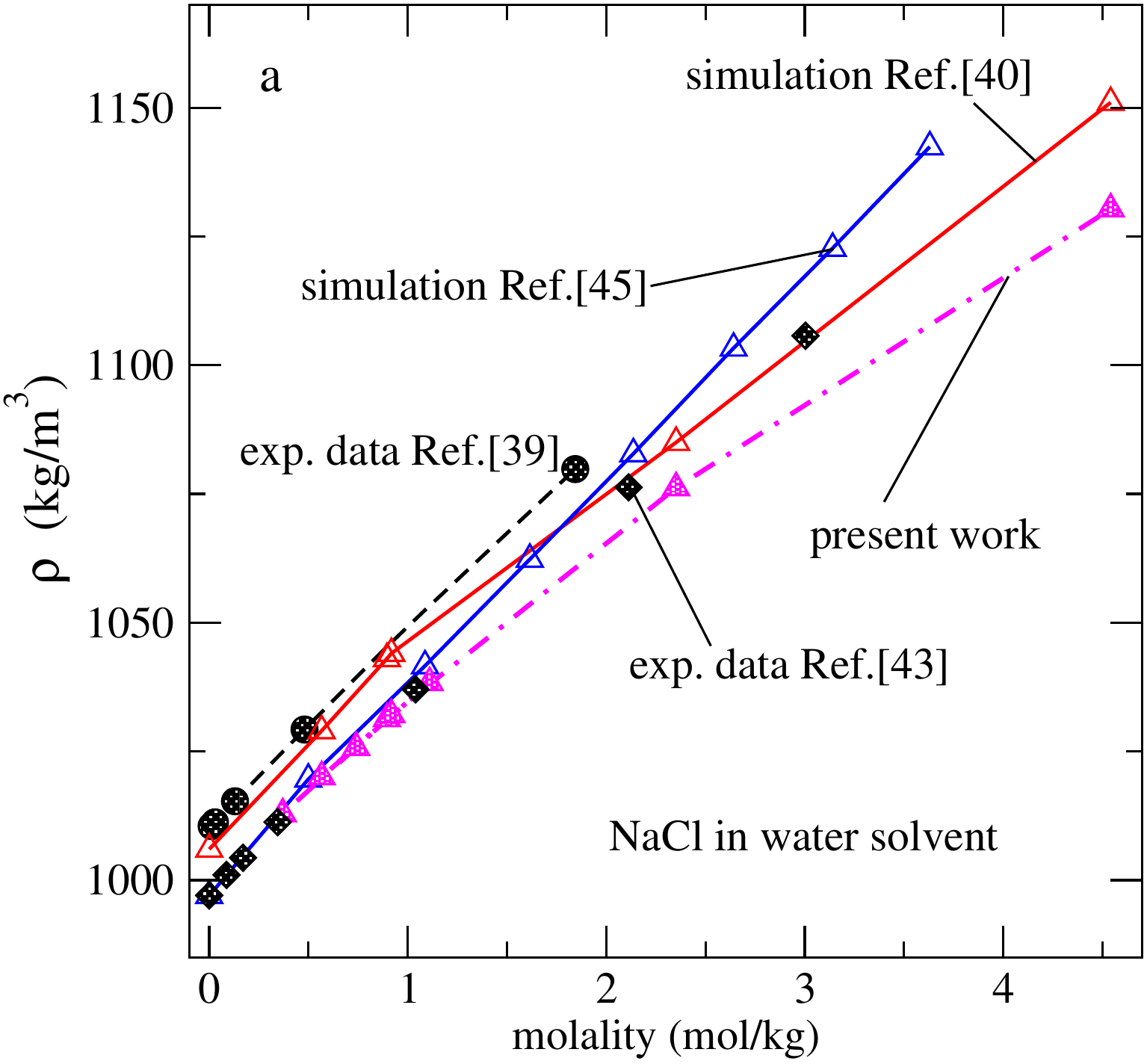} \qquad
\includegraphics[width=0.465\textwidth]{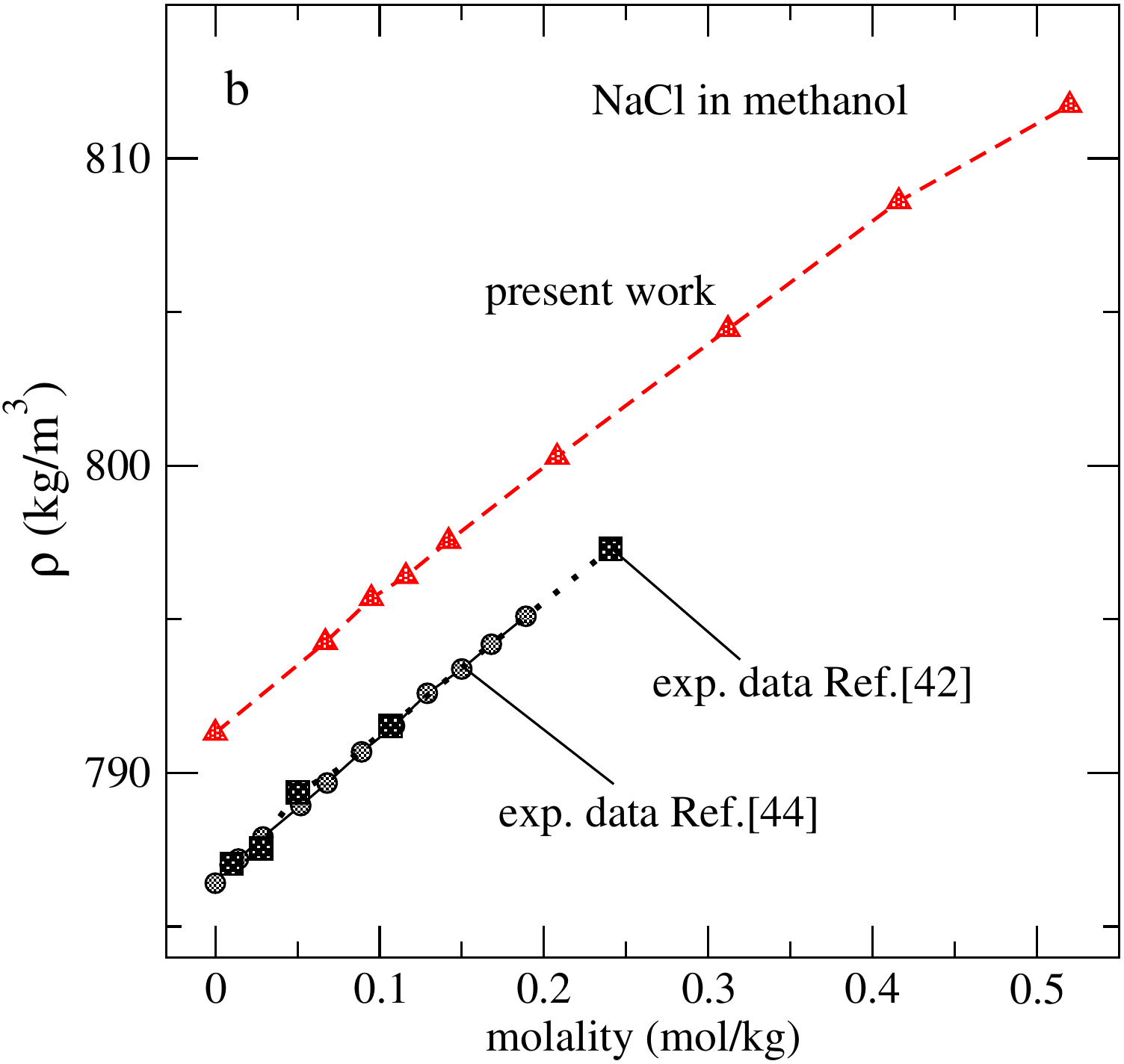}
\caption{\label{fig1} (Colour online) Panel~(a): dependence of the density of NaCl
solutions in SPC/E water on ion concentration
from constant pressure--constant temperature simulations ($T = 298.15$~K, $P = 1$~bar)
of the present work (triangles). Simulation data marked by triangles and lines 
(blue and red) are
taken from Kohns et al.~\cite{hasse} and from \cite{laaksonen}, respectively.
The experimental results marked by circles are from \cite{takenaka}. 
The experimental data marked by diamonds are from \cite{guetachew}.
Panel~(b): dependence of the density of NaCl solutions in methanol solvent (united atom model)
on ion concentration from simulations of the present work. 
Experimental data are from \cite{eliseeva} and~\cite{lankford}, circles and squares, 
respectively.} 
\end{figure}
\begin{figure}[!t]
\centering
\includegraphics[width=0.47\textwidth]{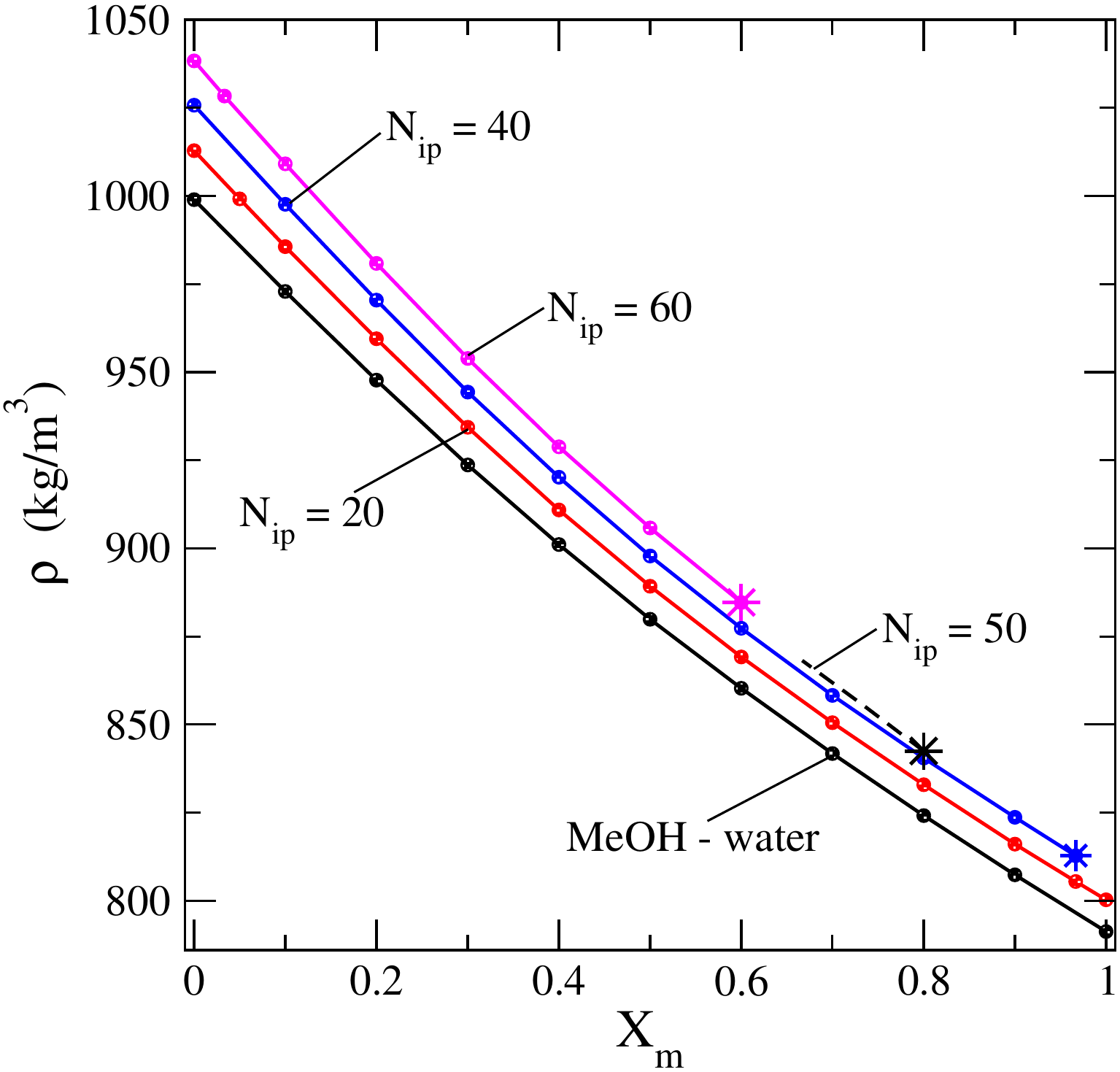}
\caption{\label{fig2} (Colour online) 
Dependence of the density of NaCl
solutions in methanol-water composite solvent on solvent composition
from constant pressure--constant temperature simulations of the 
present work ($T = 298.15$~K, $P = 1$~bar).
The lines correspond to salt-free solvent and at a different constant number of solute
molecules. The lines at $N_{\text{ip}} = 40$,  $N_{\text{ip}} = 50$ (only a fragment of the curve is plotted) 
and $N_{\text{ip}} = 60$ are given up to the 
thermodynamic conditions at which a big cluster of ions is formed. These terminating
points are marked by a star on each of the corresponding curves.}
\end{figure}

A further confirmation of this behaviour is illustrated in the left-hand panel in figure~\ref{fig1}. Both the model 
of the present study and the model used by Kohns et al.~\cite{hasse} yield a very good 
agreement with the experimental results of \cite{lankford,guetachew} at low and intermediate
molalities. At high values of molality, our model underestimates the density of the solution whereas
the model by Kohns et al.~\cite{hasse} overestimates the density. Finally, the model employed
in \cite{laaksonen} is in agreement with the experimental data of~\cite{takenaka} at
low molality and works reasonably well at intermediate values of molality. 
In the right-hand panel in figure~\ref{fig1}, we use the experimental data 
from \cite{lankford,eliseeva} for the methanolic solutions of NaCl. Our simulations 
correctly reproduce the changes of the density of these solutions depending on molality. The model
in question slightly overestimates the density but the discrepancy between simulations and 
experimental results is of the order of a fraction of one percent of the value for the density.
Having this satisfactory description of the solutions in two limiting cases, we have 
undertaken simulations for NaCl in mixed solvents, figure~\ref{fig2}. Here, we show the results for
the salt-free case and for three sets of systems with a varying solvent chemical composition.
Each set corresponds to a different fixed number of ion pairs, $N_{\text{ip}}=N_{\text{Na}}=N_{\text{Cl}}$, 
added to the solvent, namely, with $N_{\text{ip}}=20$,  $N_{\text{ip}}=40$ and  $N_{\text{ip}}=60$. 
In addition, we reproduced
a piece of the curve for a set of systems with  $N_{\text{ip}}=50$. In all cases, the number of
solvent molecules is fixed at 3000, i.e., $N_{\text w}+N_{\text m}=3000$.
The composition dependence of the density of the solution for systems with  $N_{\text{ip}}=20$
 is similar to the set of salt-free systems for all $X_{\text m}$. Interestingly, the ``insertion'' of
this amount of ion pairs into a mixed solvent results in a uniformly  augmenting density
for all $X_{\text m}$, including two limiting cases, $X_{\text m}=0$  and $X_{\text m}=1$. A further increase of the
amount of cations and anions in the system ($N_{\text{ip}}=40$,  $N_{\text{ip}}=50$ and  $N_{\text{ip}}=60$) 
leads to higher values of the density of the solution for various compositions in terms of $X_{\text m}$.
However, each of these three lines has a terminating point at which the systems reach 
saturation. Three stars in figure~\ref{fig2}, if joined together,  describe a part of the 
saturation curve. The star upon each line
describes the ``last'' homogeneous system prior to the formation of a single rather big cluster 
(involving the majority of ions, which permits to consider it as a nucleus or a grain of a solid crystal 
phase) 
at a higher value of $X_{\text m}$. This cluster coexists with a solution, i.e., with the entire solvent 
subsystem containing much less ions that are either solvated cations and anions or ion pairs.
We will return to this issue in the body of the manuscript below.

\subsection{Pair distribution functions, coordination numbers and hydrogen bonding analysis}

The microscopic structure of the solutions under study is described in terms of various
pair distribution functions (pdfs). Many of them were described in every detail 
in several publications 
for the aqueous solutions of NaCl ($X_{\text m}=0$), see, e.g.~\cite{laaksonen},  as well
as for NaCl solutions with a composite water-methanol solvent, 
see, e.g.~\cite{ewa2,ewa3,ewa4,ewa5,bouazizi}. To avoid unnecessary repetitions, we
refer to the most comprehensive and to the most recent description of the pdfs given in
\cite{bouazizi}. Moreover, the discussion by these authors involves comparisons
with previous developments and with a few experimental findings concerning a microscopic
structure of salt-free mixtures and some solutions.

\begin{figure}[!b]
\centering
\includegraphics[width=0.47\textwidth]{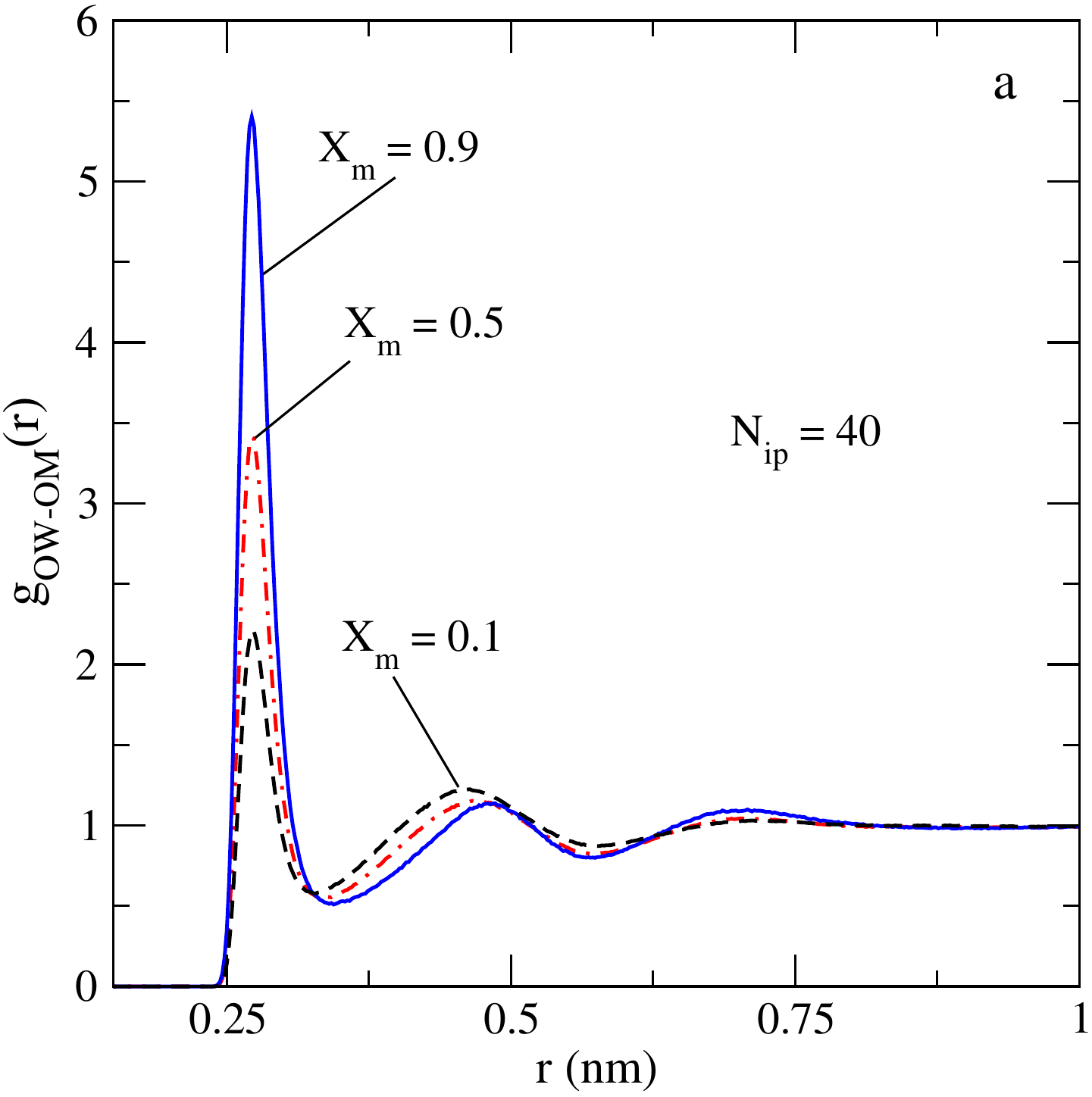} \qquad
\includegraphics[width=0.47\textwidth]{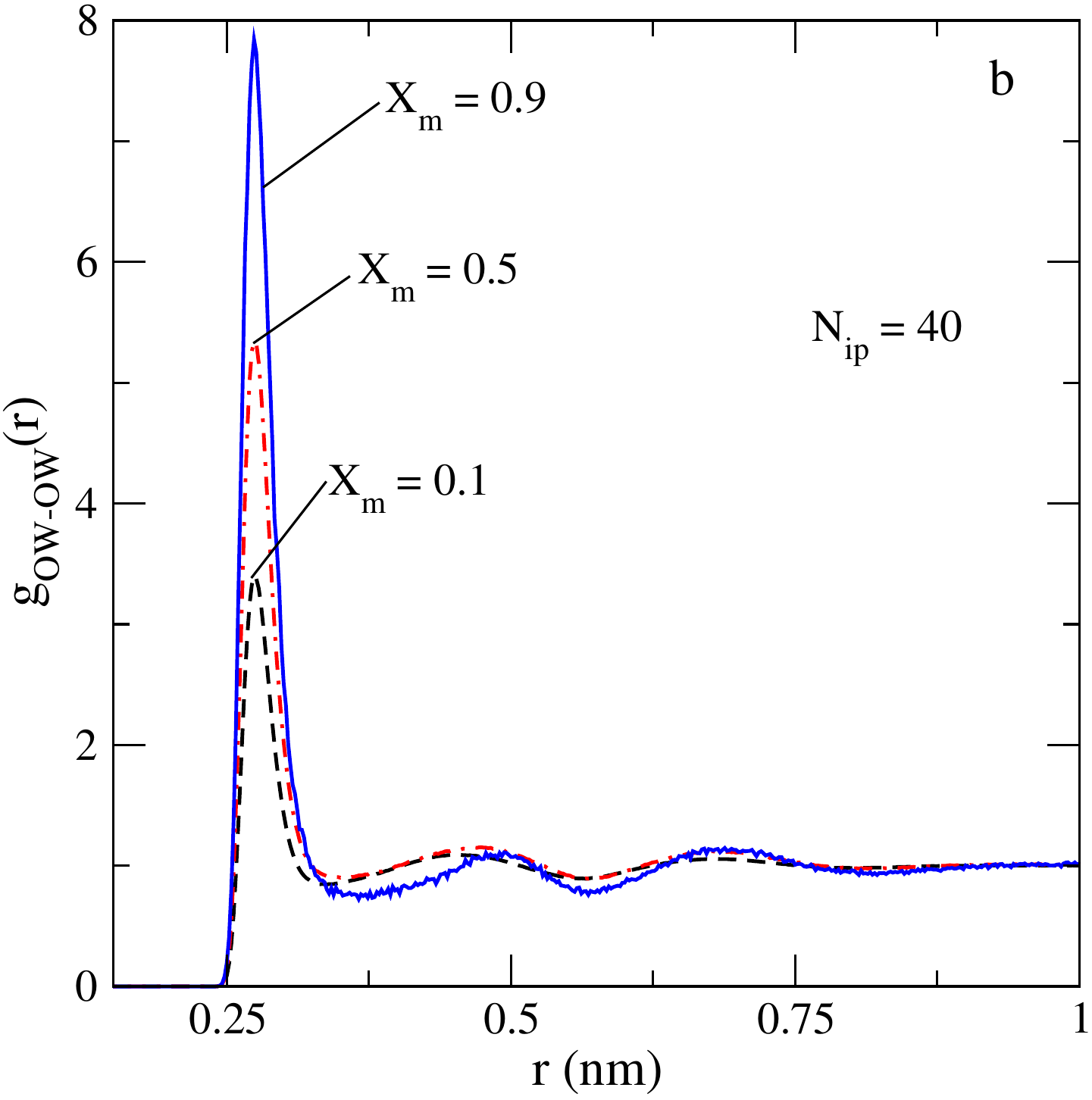}
\caption{\label{fig3} (Colour online) Pair distribution functions OW-OM and OW-OW
with changing the solvent composition.}
\protect
\end{figure}
\begin{figure}[!t]
\centering
\includegraphics[width=0.47\textwidth]{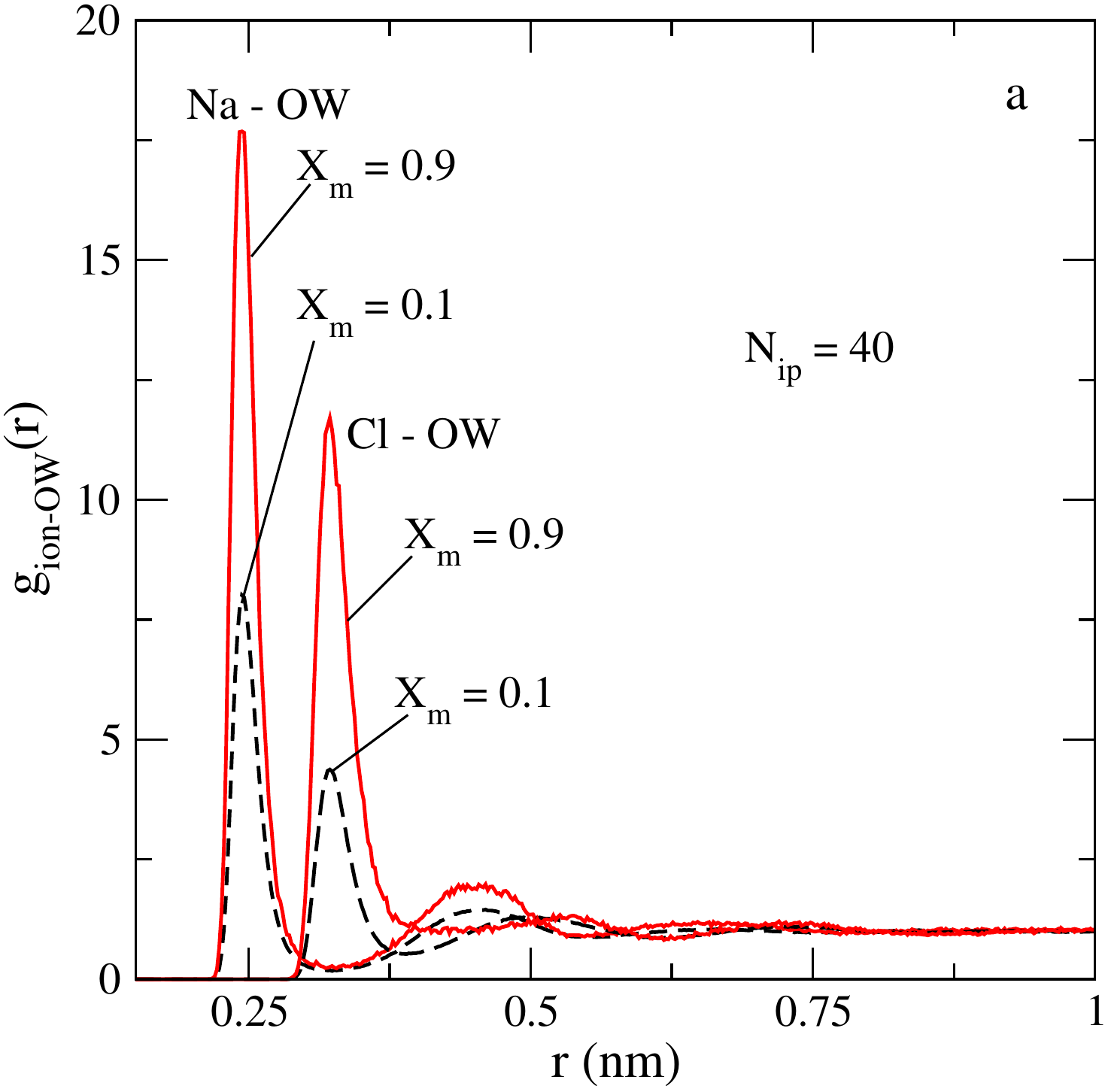} \qquad
\includegraphics[width=0.47\textwidth]{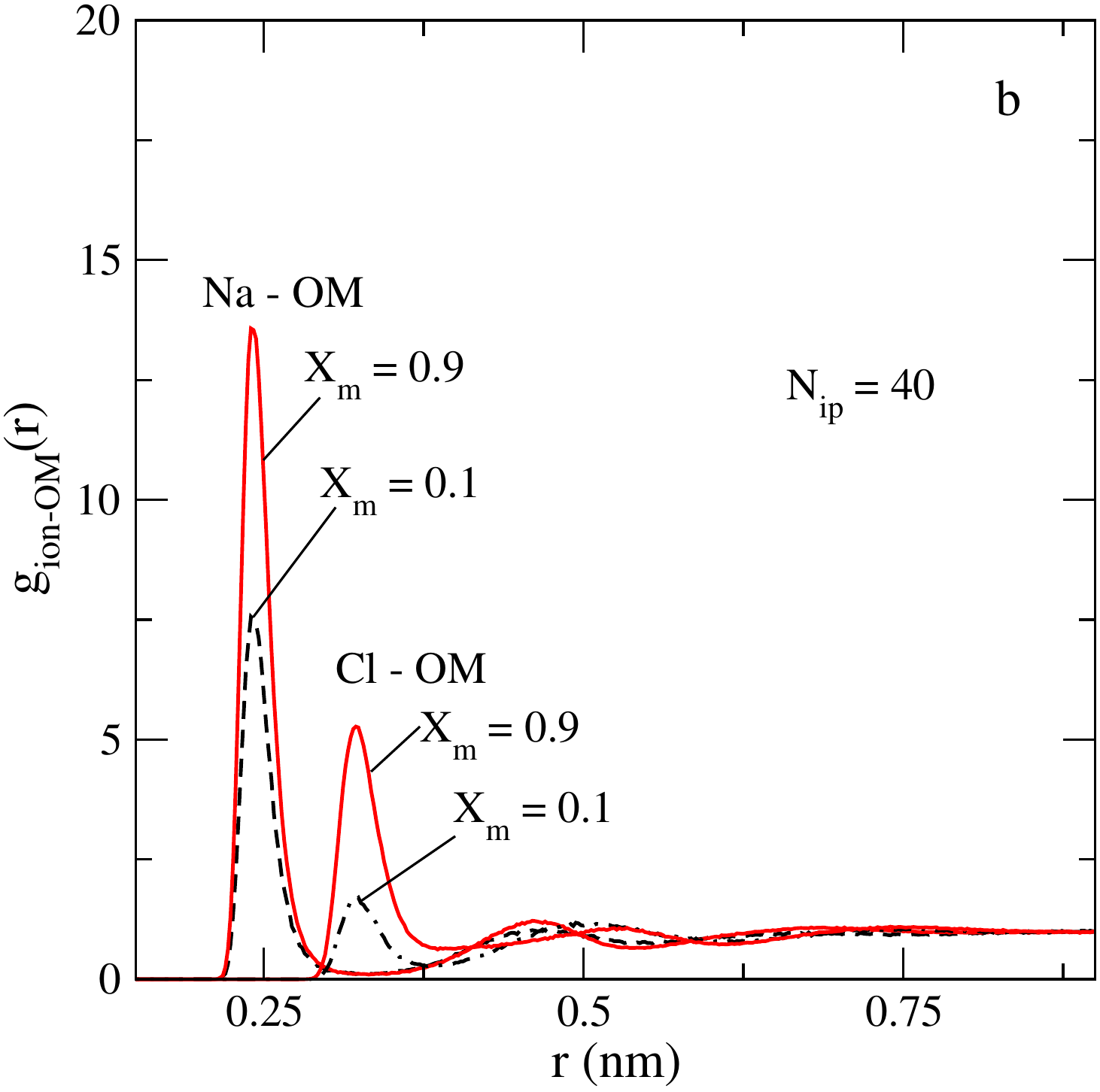}
\caption{\label{fig4}(Colour online) Pair distribution functions Na-OW and Cl-OW (panel a)
and Na-OM and Cl-OM (panel b) with changing the solvent composition.}
\protect
\end{figure}

Our few examples, shown in figures~\ref{fig3} and \ref{fig4}, just serve to confirm that
the model of the present study leads to correct and physically well grounded
conclusions in accordance with previous developments. The results are given
solely for a single set of systems, namely the number of solutes is fixed at $N_{\text{ip}}=40$
while the solvent composition changes with $X_{\text m}$, the total number of solvent
molecules is kept fixed at $N_{\text w}+N_{\text m}=3000$.
As concerns the behaviour of the cross pdfs between solvent species, $g_{\text{OW-OM}}(r)$ in figure~\ref{fig3}, 
we observe that the first maximum gradually grows whereas the second maximum monotonously
decreases with an increasing amount of methanol in the solvent. Thus, the water-methanol
contacts become more probable with an increasing $X_{\text m}$. However, the correlations between
two species at a larger scale become weaker. This behaviour is in
accordance with the trends observed for the salt-free systems. 
At the same time, the short-range correlations within a water subsystem become stronger
with an increasing $X_{\text m}$, as evidenced by the growth of the first maximum of
$g_{\text{OW-OW}}(r)$ shown in the right-hand panel in figure~\ref{fig3}. The positions of the first maxima
describing most probable configurations of solvent species 
in terms of $g_{\text{OW-OM}}(r)$ and $g_{\text{OW-OW}}(r)$ are practically unchanged with the changes of $X_{\text m}$.
Most probably, the correlations between more distant molecules are affected by changes
of the solvent chemical composition, see~\cite{bouazizi} for an ampler discussion. 

As concerns the behaviour of ion-water pdfs, we would like to mention the following. 
In the water-rich solvents, see, e.g. the curves corresponding to $g_{\text{ion-OW}}(r)$ at
$X_{\text m}=0.1$, ions are very well distributed in the solvent, figure~\ref{fig4} (left-hand panel).
This is evidenced by a high first maximum of $g_{\text{Na-OW}}(r)$ and a well pronounced
second maximum of this function. On the other hand, the height of the first maximum of
$g_{\text{Cl-OW}}(r)$ and of $g_{\text{OW-OW}}(r)$ are nearly equal. Inspection of the 
curves describing $g_{\text{Na-OM}}(r)$ and $g_{\text{Cl-OM}}(r)$ (right-hand panel in figure~\ref{fig4}) 
and their comparison with  the $g_{\text{Na-OW}}(r)$ and $g_{\text{Cl-OW}}(r)$ functions 
leads to the following conclusions. The methanol molecules are capable of approaching the
sodium cation as satisfactorily as the water molecules do, in spite of the number of methanol
particles being quite low at $X_{\text m}=0.1$. On the other hand, the chloride anion is
preferentially surrounded by waters, the methanol species (OM) possibly
enter the solvation shell of Cl$^-$, though with much lower probability compared to OW.
In solvents that are rich in methanol ($X_{\text m}=0.9$), we observe that sodium cation 
attracts as much waters as available, but the solvation shell contains some methanol 
molecules since the first maximum of $g_{\text{Na-OM}}(r)$ is quite high.
The chloride anions also attract as many water molecules as available at this solvent
composition, but the magnitude of the first maximum of the function $g_{\text{Cl-OM}}(r)$ substantially
grows upon the changes of $X_{\text m}$ from 0.1 to 0.9.

\begin{figure}[!b]
\centering
\includegraphics[width=0.47\textwidth]{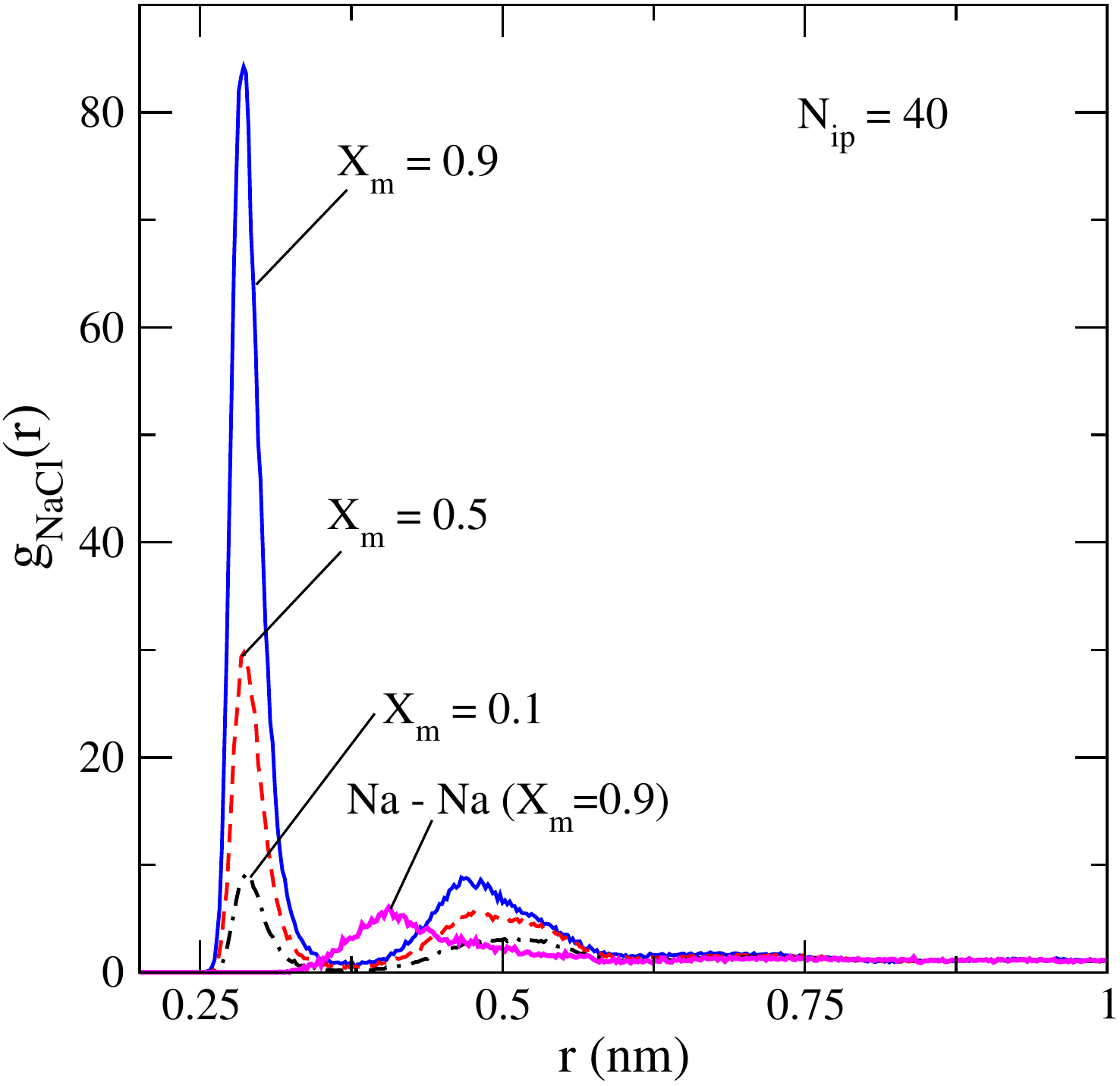}
\caption{\label{fig5}(Colour online) Pair distribution functions
and  Na-Cl with changing the solvent composition.}
\end{figure}

The ion-ion pair distribution function, $g_{\text{Na-Cl}}(r)$ (in  figure~\ref{fig5}) 
exhibits a maximum at $r \approx 2.8$~{\AA} describing the probability of finding 
contact ion pairs, the
second maximum at $r \approx 5$~{\AA} showing the  presence of solvent-shared 
ion pairs, while the following oscillations are not visible on the scale of the figure.
The first and the second maxima essentially increase in magnitude with 
the change of the solvent composition. 
At the same time, the pdfs describing the configurations of similarly charged ions are 
characterized by a more pronounced structure [we show the function
$g_{\text{Na-Na}}(r)$ at $X_{\text m}=0.9$]. It is worth mentioning that in the case of dilute 
solutions, the trends of the behaviour of the ion-ion pdfs with ion concentration are in 
accordance with the calculations of the potential of the mean force, see, e.g.~\cite{vlachy}.
In the present case, at $X_{\text m}=0.9$, we observe the development of anti-phase 
oscillations that witness the formation of short-range ordering of ions, 
which results in cluster formation upon a further increase of the methanol content.

We should like to finish the discussion of the pair distribution functions with 
the following comment.
We believe that the model yields a qualitatively correct and a physically sound picture, 
while other combinations of the force fields yield formally similar results. 
However, any kind of improvement could be attempted only if the experimental 
structure factors of such complex systems were  available.
Then, the fitting of computer simulation results to, e.g., diffraction experiment 
data, in spite of the expected technical difficulties, could be performed 
along the lines proposed in \cite{pusztai2,pusztai3} for aqueous 
electrolyte solutions.

The pair distribution functions yield the running coordination numbers through the 
equation,
\begin{equation}
 n_{ij}(R)=4\piup\rho_j\int_{0}^{R}g_{ij}(r)r^{2}\rd r,
% \label{eq_nr}
\end{equation}
where $\rho_j$ is the number density of species $j$. The first coordination number 
is obtained  by putting $R=r_{\text{min}}$, i.e., at the first minimum of the
corresponding pair distribution function. 

\begin{figure}[!b]
\centering
\includegraphics[width=0.47\textwidth]{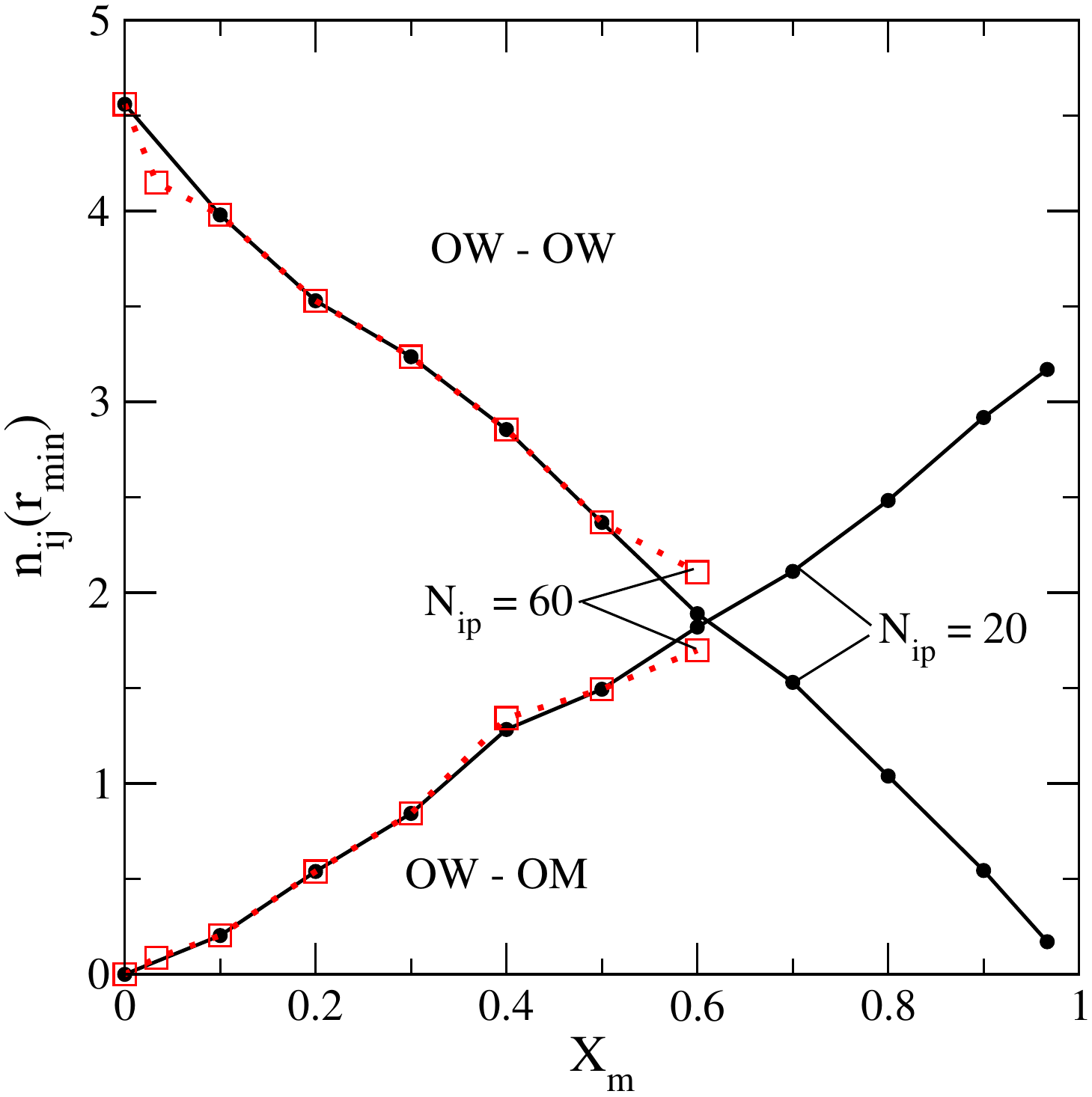}
\caption{\label{fig6}(Colour online) Changes of the first coordination numbers, $n_{\text{OW-OW}}$ 
and $n_{\text{OW-OM}}$, upon changing the solvent composition, $X_{\text m}$, for systems with
$N_{\text{ip}}=20$ and $N_{\text{ip}}=60$.}
\protect
\end{figure}

\looseness=-1 As concerns the coordination numbers describing the solvent species, we can observe that
 $n_{\text{OW-OW}}$ coordination number almost monotonously decreases if $X_{\text m}$ grows
from zero to unity, figure~\ref{fig6}. It starts at a value $\approx 4.5$ close to what
is  expected for water.  At the same time, $n_{\text{OW-OM}}$ increases.
The corresponding line ends up at $\approx 3.2$, indicating a less coordinated water molecule
by methanol particles in comparison with the aqueous medium.
We were unable to detect pronounced overall differences between the coordination numbers of
solvent species upon changes of the number of solute ions from $N_{\text{ip}}=20$ to
$N_{\text{ip}}=60$. Just in the latter case the lines deviate from 
the results corresponding to $N_{\text{ip}}=20$  and approach the saturation conditions
at $X_{\text m} \approx 0.6$. The shape of the lines joining the calculated points is approximately  
linear.

\begin{figure}[!t]
\centering
\includegraphics[width=0.465\textwidth]{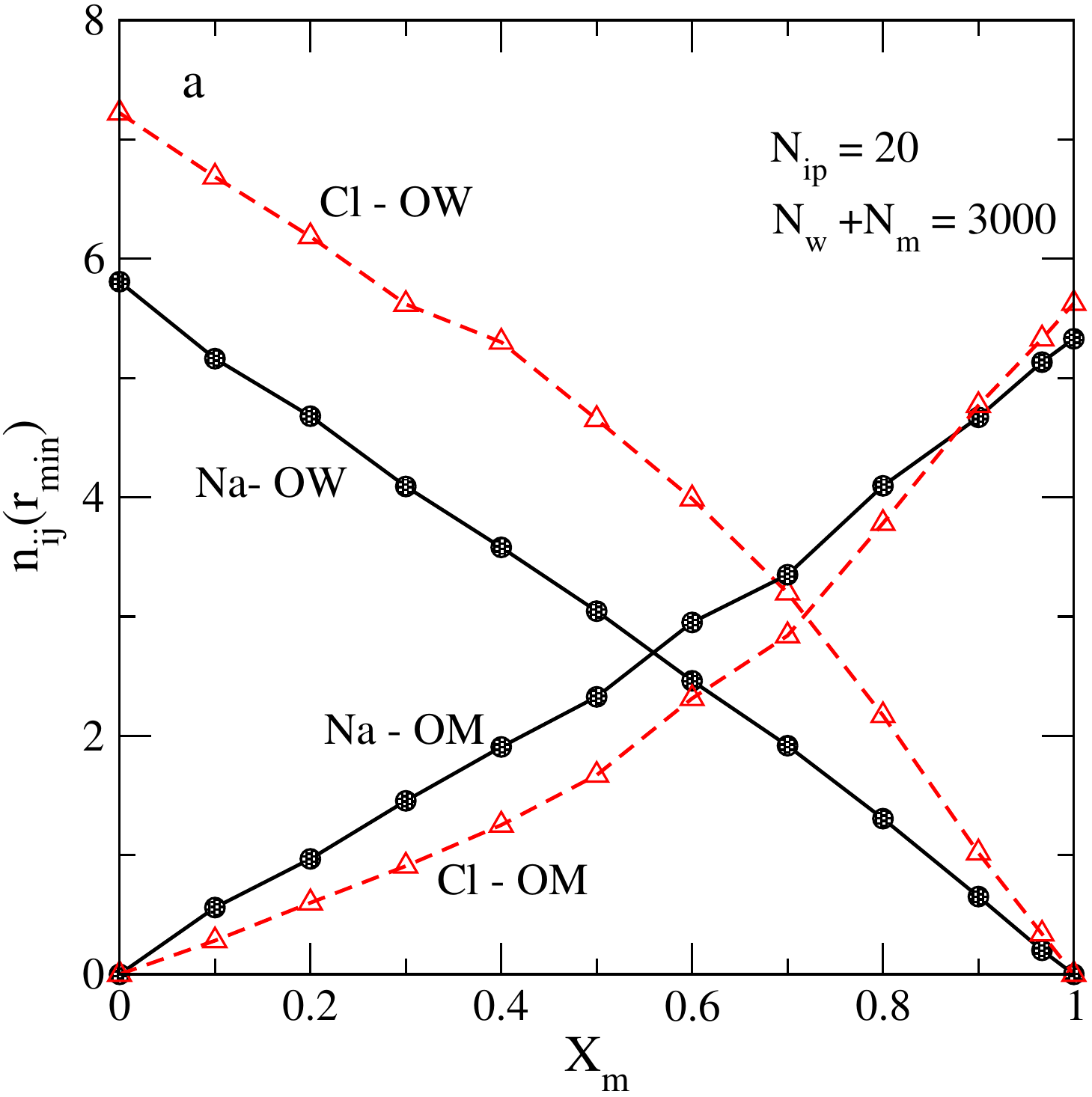} \qquad
\includegraphics[width=0.465\textwidth]{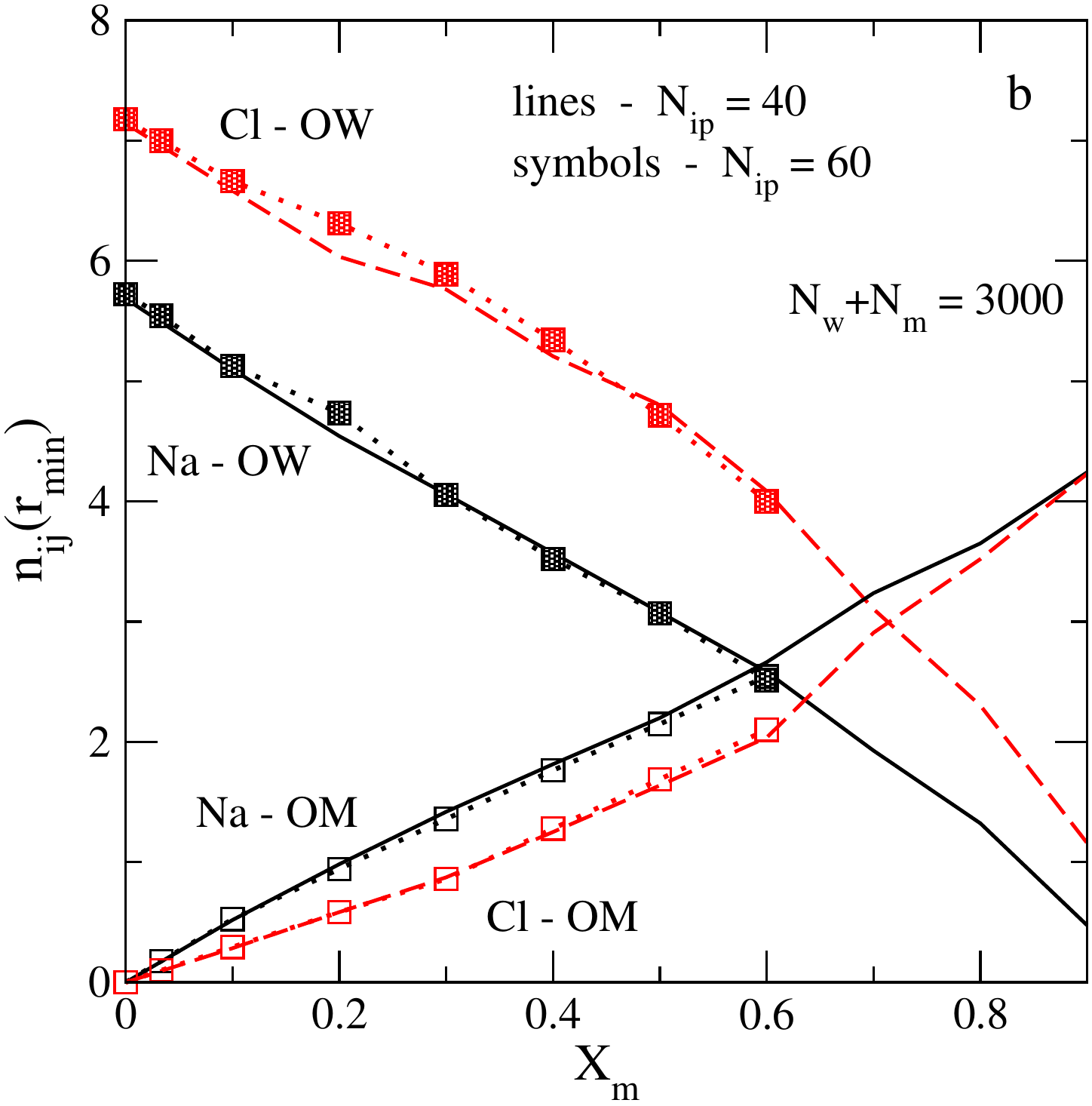}
\caption{\label{fig7}(Colour online)
Changes of the first coordination number of
cations and anions, $n_{\text{Na-}i}$  and $n_{\text{Cl-}i}$  ($i=\text{OW}, \text{OM}$), with
changing the solvent composition.}
\protect
\end{figure}

The behaviour of the coordination numbers of ions and solvent species exhibits 
certain interesting features and peculiarities. Namely, from the left-hand panel in figure~\ref{fig7}
($N_{\text{ip}}=20$),
we learn that $n_{\text{Na-OW}}$ linearly decreases with increasing $X_{\text m}$ whereas
the coordination number $n_{\text{Na-OM}}$ almost linearly increases. The coordination number
$n_{\text{Cl-OW}}$ decreases as well, upon increasing $X_{\text m}$. However, it follows the trends of
$n_{\text{Na-OW}}$ until $X_{\text m} \approx 0.4$. In the interval $X_{\text m} > 0.6$, the decay of  $n_{\text{Cl-OW}}$ 
is much faster. A qualitatively similar shape with two different rates of changes 
is seen in $n_{\text{Cl-OM}}$. The difference in coordination of Na and Cl at $X_{\text m}=0$ is
in accordance with the previous studies of aqueous solutions. However, we can observe that
the difference of coordination of sodium cation in pure water and in pure methanol
is not big. On the other hand, the chloride anion is much less coordinated in methanol
compared to water. 

The values for the coordination numbers at $X_{\text m}=1$ in the left-hand
panel in figure~\ref{fig7} agree with the results obtained recently in \cite{reiser}, 
see table~9 of that reference. Similar trends of behaviour of ion-solvent species
coordination numbers are observed for a higher number of solute ions, $N_{\text{ip}}=40$.
While approaching saturation conditions, the coordination numbers of cations and 
anions in the case $N_{\text{ip}}=40$ become close to each other, resembling what happens in the 
systems with a lower number of solutes in the solvent with a predominating methanol
amount. However, the system with $N_{\text{ip}}=60$ can be very close to saturation, but
the coordination numbers of sodium cations and of chloride anion differ much
(see dotted curves with squares in the right-hand panel in figure~\ref{fig7}). Thus, the ion-solvent
coordination numbers do not indicate on their own that the system approaches 
saturation.

\begin{figure}[!t]
\centering
\includegraphics[width=0.47\textwidth]{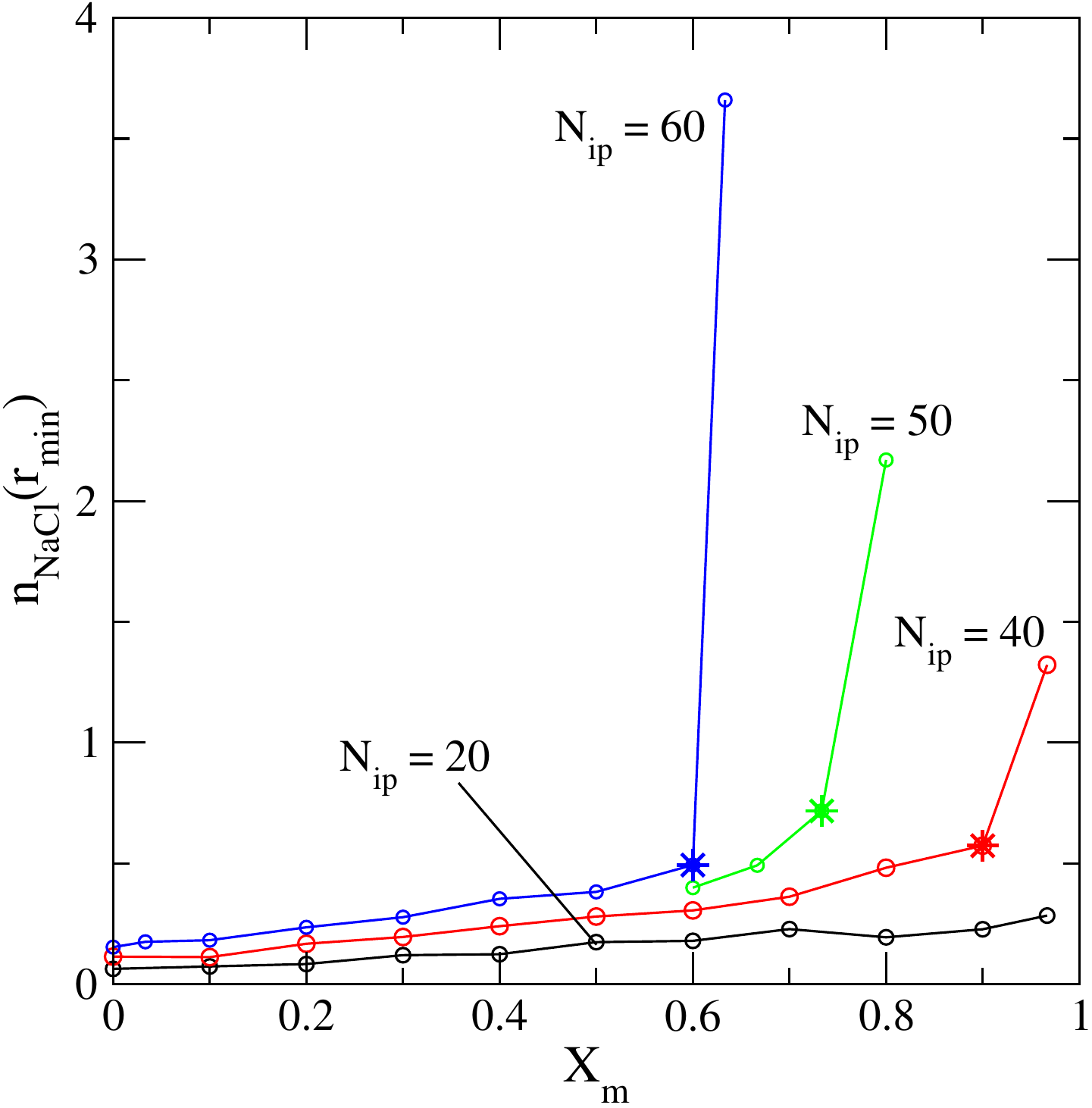}
\caption{\label{fig8}(Colour online) Changes of the first coordination number $n_{\text{Na-CL}}$ with
changing the solvent composition at three ion concentrations studied.}
\protect
\end{figure}

Finally, the first coordination number, $n_{\text{Na-Cl}}$,
as a function of the solvent composition is given in figure~\ref{fig8} for three sets of 
systems differing by the ion
fraction. All the curves behave qualitatively similarly: the 
coordination number grows with increasing $X_{\text m}$ and exhibits a substantial jump 
at conditions when a rather  big ionic cluster is formed.
The magnitude of the jump depends on the number of solute molecules as indicated in
figure~\ref{fig8}. Thus, the  $n_{\text{Na-Cl}}$ exhibits a single and well pronounced peculiarity in a water-methanol
solvent. 

The threshold value for the observed jump depends on the solvent composition (one may express
it as $X_{\text m}$ or in terms of weight percentage of the co-solvent) and on the ion concentration
(usually molarity or molality scale is used in experimental studies). Certainly,
the precise values for threshold conditions depend on the details of
the force field of ions as well as on the type of the models for water and methanol,
e.g., united atom versus all-atom modelling. Therefore, we postpone a more detailed
exploration of this issue in terms of the coordination numbers to a future work.
However, it is worth mentioning that the threshold situation can be monitored by the behaviour
of the self-diffusion coefficients of ions and by the conductivity of the solutions,
besides the coordination numbers.

\begin{figure}[!t]
\centering
\includegraphics[width=0.47\textwidth]{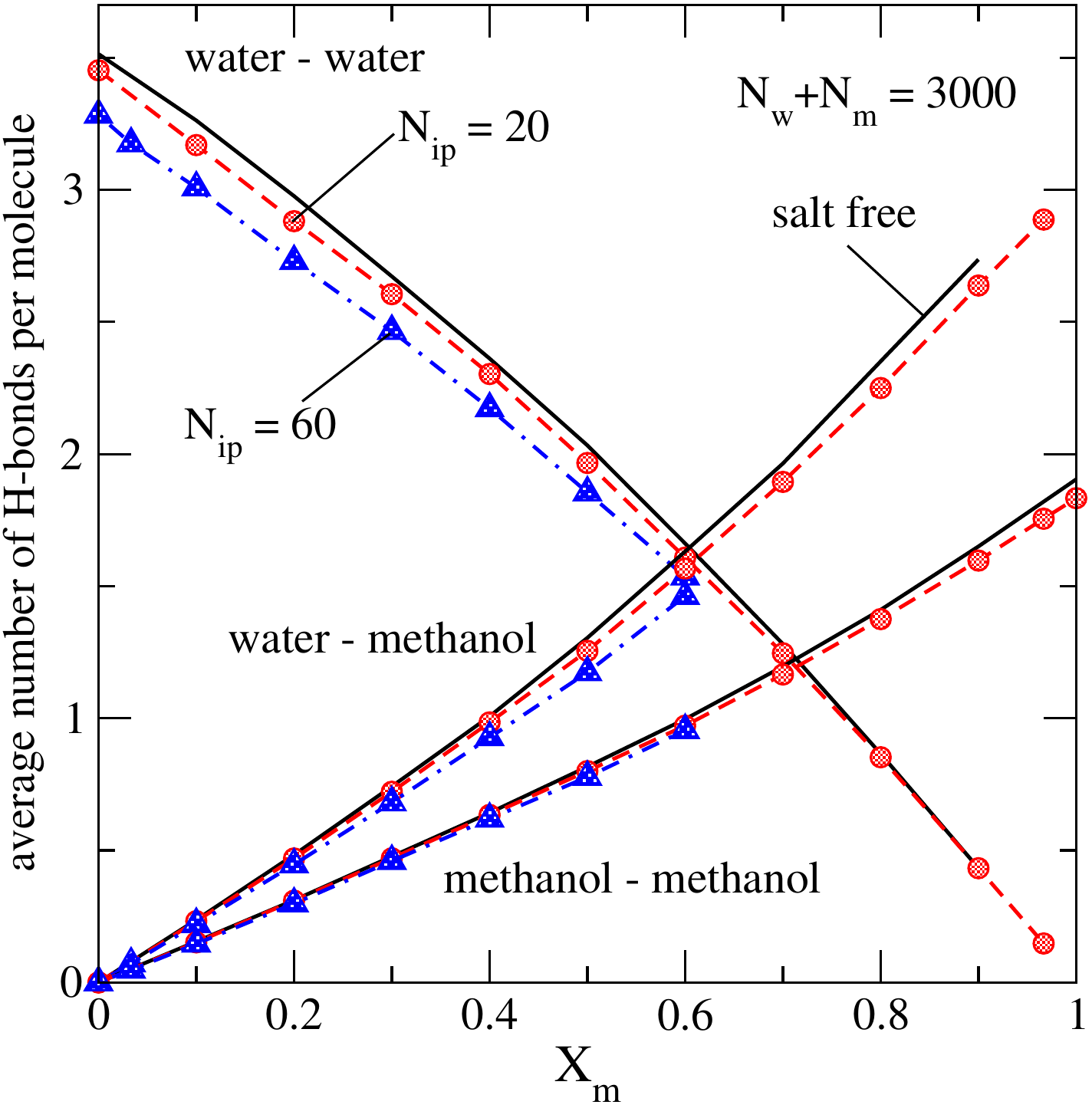}
\caption{\label{fig9}(Colour online) Average number of hydrogen bonds between solvent species upon
composition changes in the NaCl solutions with composite solvent.}
\protect
\end{figure}

Our final concern in this section is in describing the changes of hydrogen bonds in a
solvent subsystem of the solutions under study upon composition changes.
The average number of hydrogen bonds per water molecule, as a function of $X_{\text m}$, 
is shown in figure~\ref{fig9}. Technically this was obtained using the gmx hbond utility with
default options of the GROMACS software.

The curves are rather smooth and exhibit certain similarities to the behaviour of 
coordination  numbers discussed above. Furthermore, for the sake of comparison, 
we performed necessary calculations for the salt-free systems of variable 
composition. The limiting values for water ($X_{\text m}=0$) and for methanol ($X_{\text m}=1$)
are in accord with the previous study from this laboratory~\cite{galicia2}.

In the entire interval of $X_{\text m}$, the average number of H-bonds per water
molecule $\langle n_{\text{HB}}\rangle$ is lower in the system  with ions in comparison with a salt-free system 
at the same $X_{\text m}$.  The cumulative effect of solutes (cations and anions) is that they actually
make the formation of a certain amount of water-water H-bonds impossible 
 due to the existence of the solvation shells. By considering the magnitude of changes,
this cumulative effect can be referred to as primary effect.
At a fixed $X_{\text m}$, the average numbers of the cross water-methanol H-bonds 
(per the number of water molecules) in the combined solvent, i.e., in the 
absence of ions, and in the solution are close to each other.  Finally,
even a weaker effect is observed on the average number of bonds formed 
between methanol molecules. 
From the analyses above, it follows that ions affect the properties of the 
system mainly via the
modification of the structure of a water subsystem of the solvent. As a result, the trends of mixing  
of water with methanol may display certain changes but not big ones.

\begin{figure}[!t]
\centering
\includegraphics[width=0.485\textwidth]{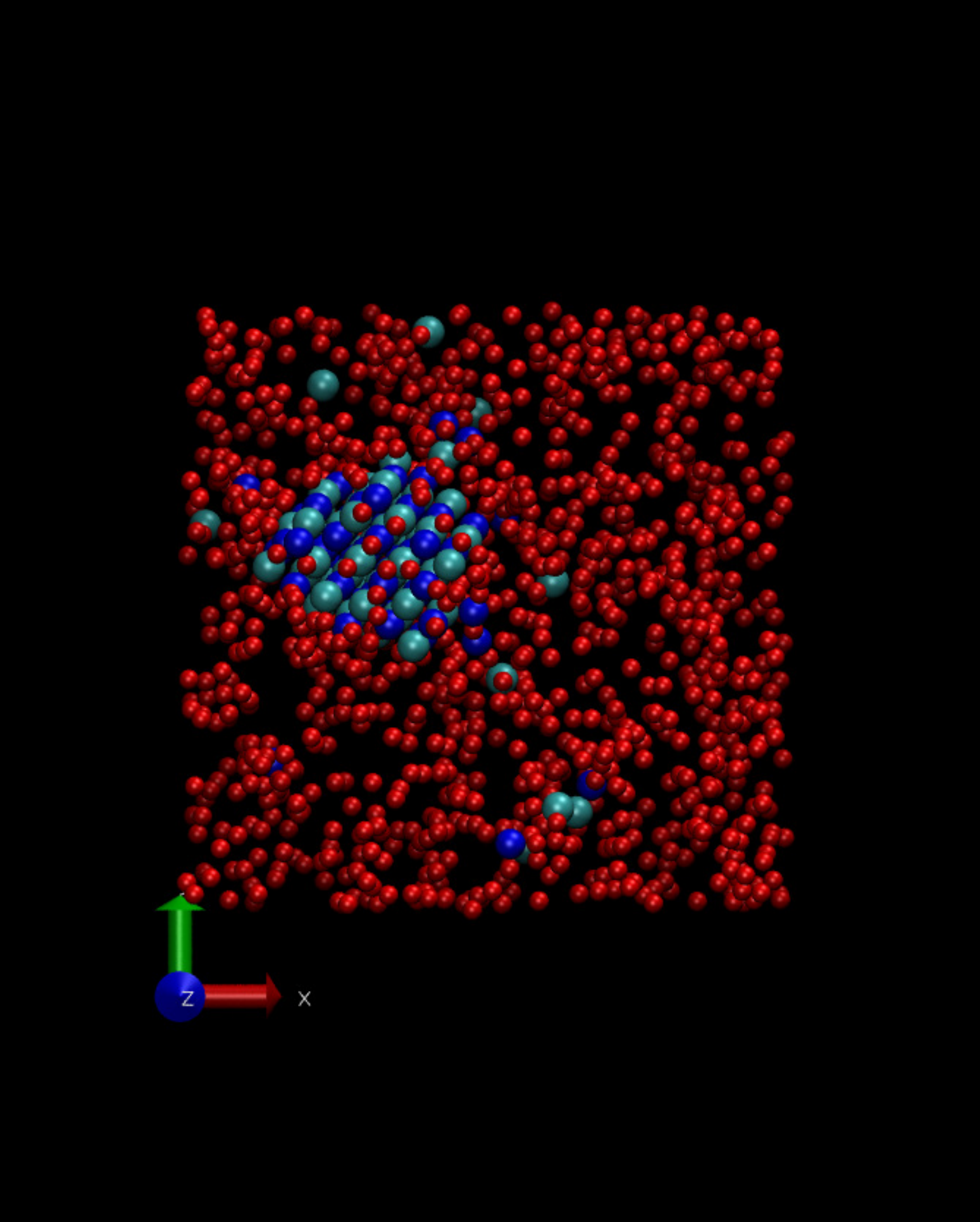}
\caption{\label{fig10}(Colour online) Snapshot describing distribution of ions (bigger spheres)
and water molecules (smaller red spheres) in the simulation box just after cluster
of cations and anions has been formed,  $X_{\text m} = 0.633$, $N_{\text{ip}}=60$, 
$N_{\text w}+N_{\text m}=3000$.}
\protect
\end{figure}

Final part of this subsection is devoted to the description of conditions under which the ion
clusters are formed.  This problem has already been considered
for NaCl aqueous solutions in several publications, 
see, e.g.,~\cite{mendoza,spohr,degreve}.
The cluster (or clusters) formed during the final part of a long trajectory obtained in
simulations is the
precursor or a small grain of the salt precipitate. The study of the solubility
of a  given salt
in a solvent of different composition is beyond the scope of the present work
at this stage of our project. Nevertheless,
we have picked up the final frame of the MD trajectory for the system with
$N_{\text{ip}}=60$, $X_{\text m}=0.6333$ ($N_{\text w}+N_{\text m}=3000$) and made visualization
of the coordinates of particles. 
It is given in figure~\ref{fig10} (water molecules in the box are shown besides
the ions). Apparently, a nice picture with a pronounced symmetry
of ions in a rather big cluster does not provide a description of the structure 
formed in quantitative terms \cite{wang}. Still, one can appreciate that not all the ions belong
to the cluster. On the other hand, in spite of macroscopic miscibility, 
heterogeneity of the distribution of solvent species on the local scale 
(say, the scale of the simulation cell) is seen, methanol molecules fill the 
``empty'' space where water particles and ions are not present. 

In order to get a better and more quantitative insight into the formation of a big
cluster of ions, we plot the number of clusters, the maximal cluster size and the average cluster
size as functions of time emerging from the trajectory of the system.
Technically, these properties result from the gmx cluster utility.
The parameter for distance to count the ions belonging to a cluster is taken 
a bit larger than the location of the first maximum of the $g_{\text{Na-Cl}}(r)$.
A bit different choice can yield slightly different numbers, but qualitatively the trends 
in the  behaviour of the properties describing clusters remain the same. 

\begin{figure}[!t]
\centering
\includegraphics[width=0.47\textwidth]{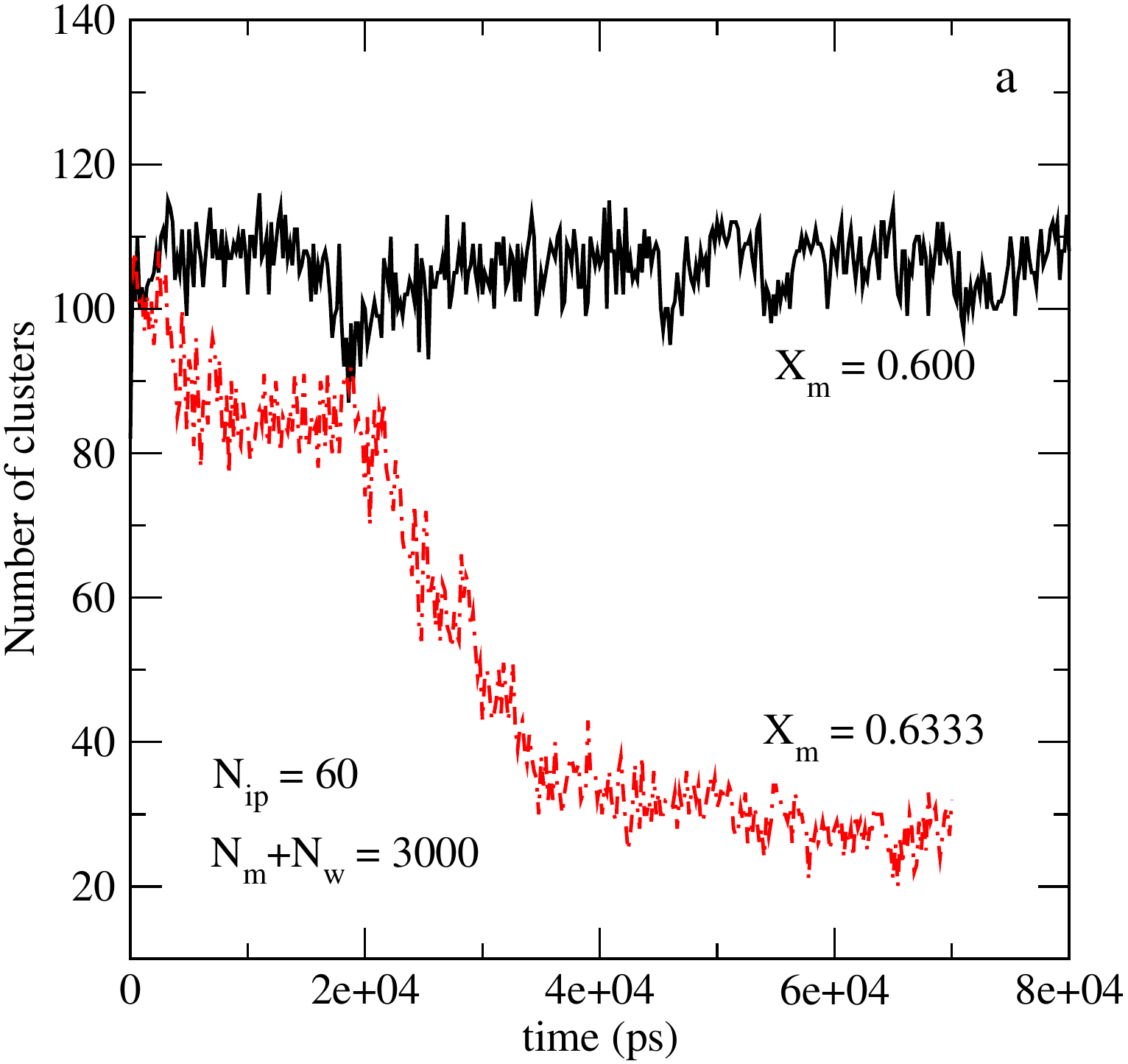} \qquad
\includegraphics[width=0.47\textwidth]{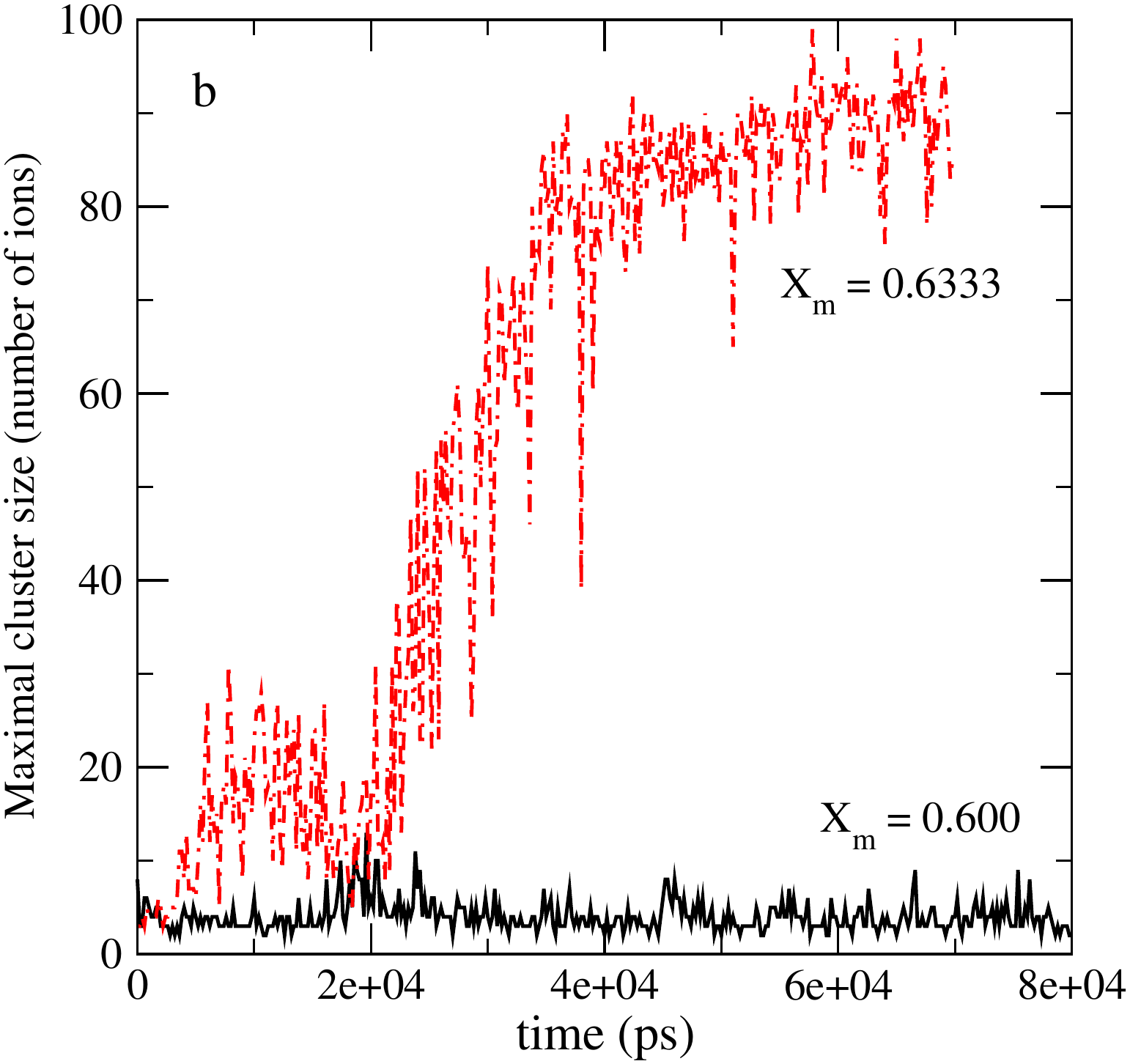}\\
\includegraphics[width=0.47\textwidth]{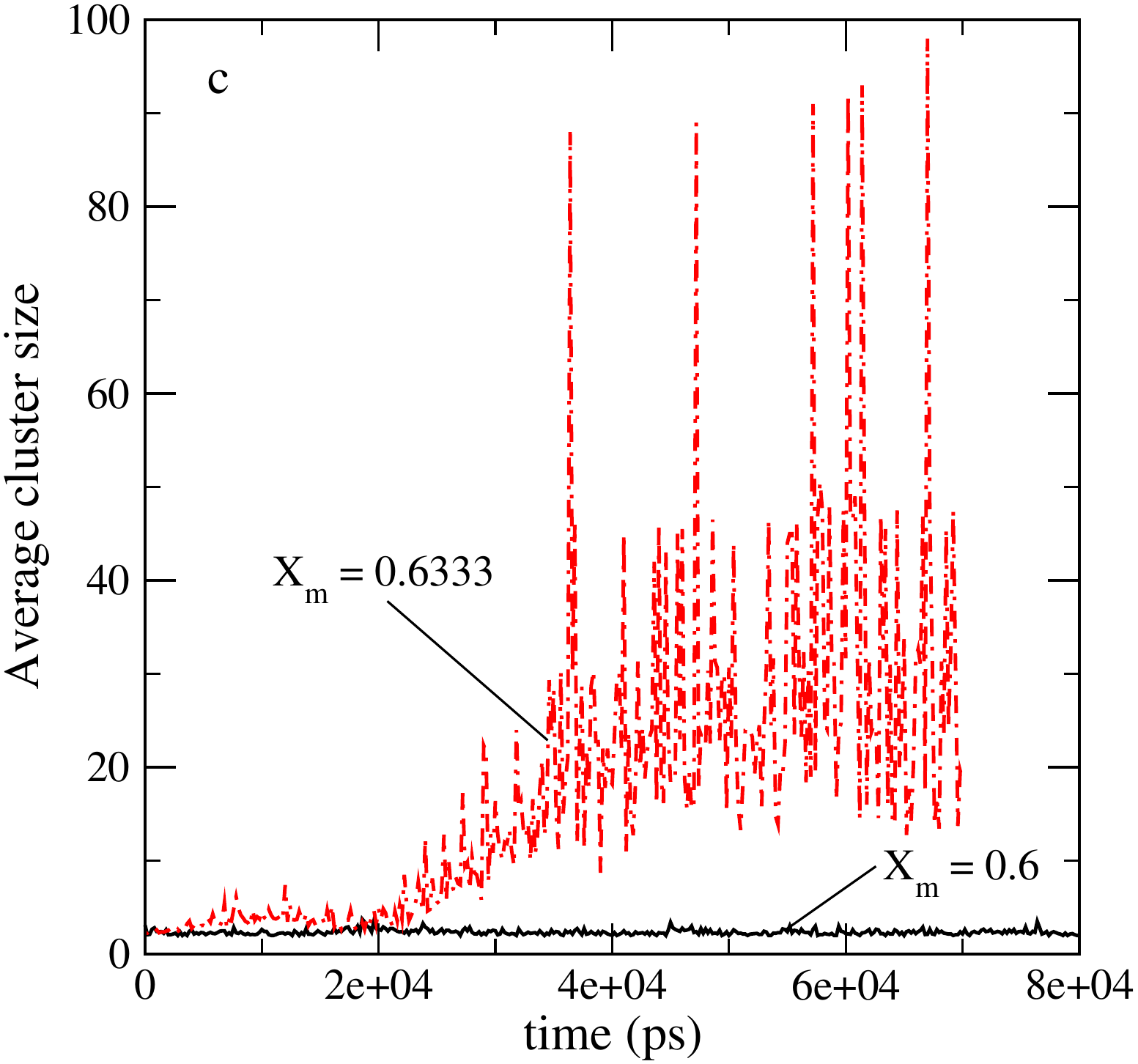}
\caption{\label{fig11}(Colour online) Panel~(a): time evolution of the number of clusters
prior to and after saturation. Panels~(b) and (c): the maximal cluster size and
the average cluster size, prior to and after crossing the saturation line, respectively.}
\protect
\end{figure}

Inspecting figure~\ref{fig11}, we can observe that prior to saturation conditions, i.e., at $X_{\text m}=0.6$,
the number of clusters fluctuate in the interval approximately between 100 and 110 
(recall that we have 120 ions). In other words, the majority of ions are ``free'' and
only some of them presumably form ion pairs along the entire trajectory. On the
other hand, crossing the saturation leads to drastic changes of this function. 
After approximately
20~ns, the number of clusters drops to reach a stable value around 30 clusters.
The maximal cluster size, in terms of the number of ions participating in it, becomes around
90, the average cluster size attains a much bigger value for $X_{\text m}=0.633$ 
in comparison with $X_{\text m}=0.6$. Thus, the majority of ions
can really  be considered as those belonging to a rather big cluster, the rest of the ions can be free or
be involved in small entities such as ion pairs, for example. 

\subsection{Self-diffusion coefficients and the dielectric constant}

\begin{figure}[!t]
\centering
\includegraphics[width=0.47\textwidth]{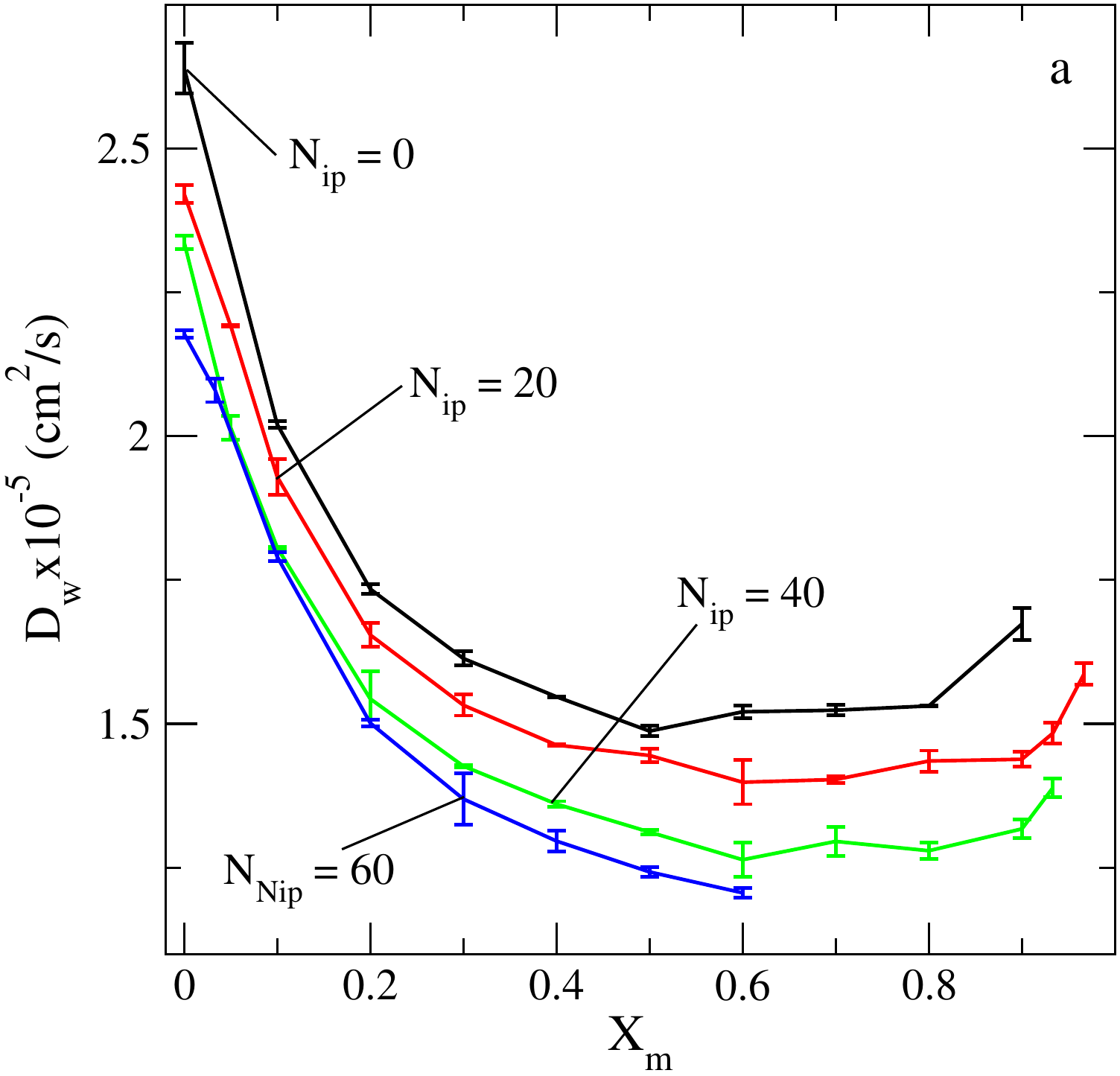}\qquad
\includegraphics[width=0.47\textwidth]{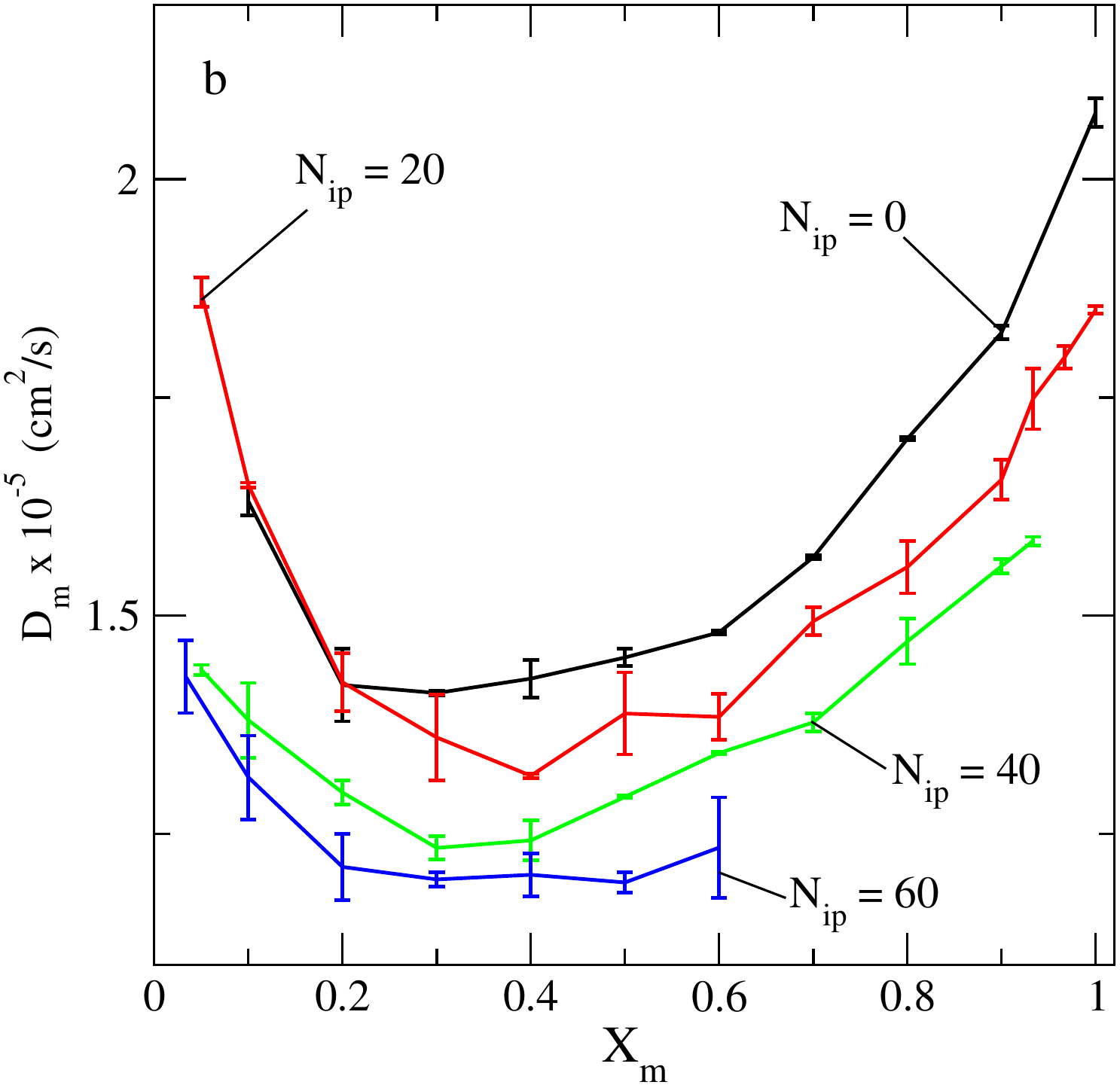}\\
\includegraphics[width=0.47\textwidth]{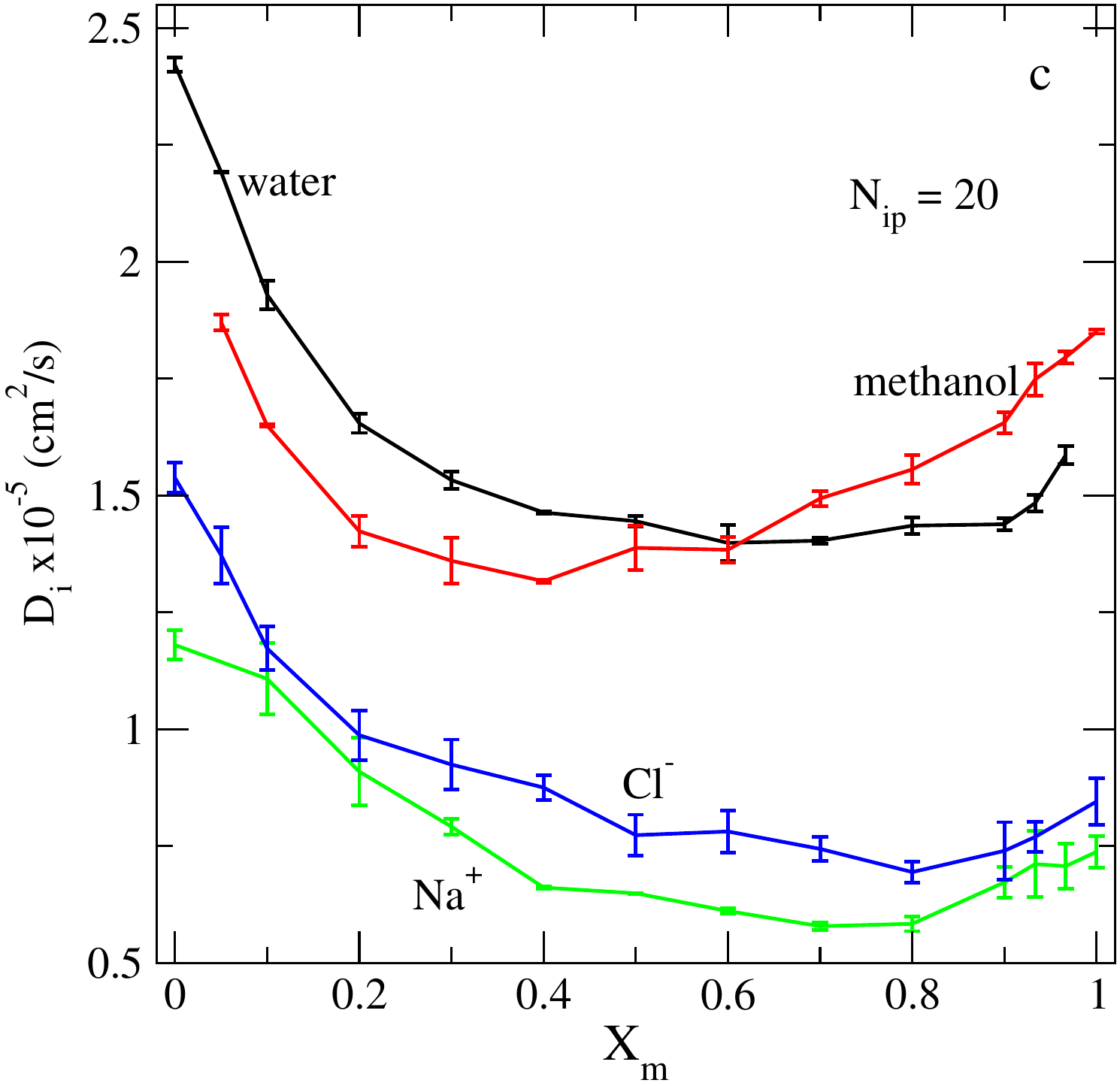}\qquad
\includegraphics[width=0.47\textwidth]{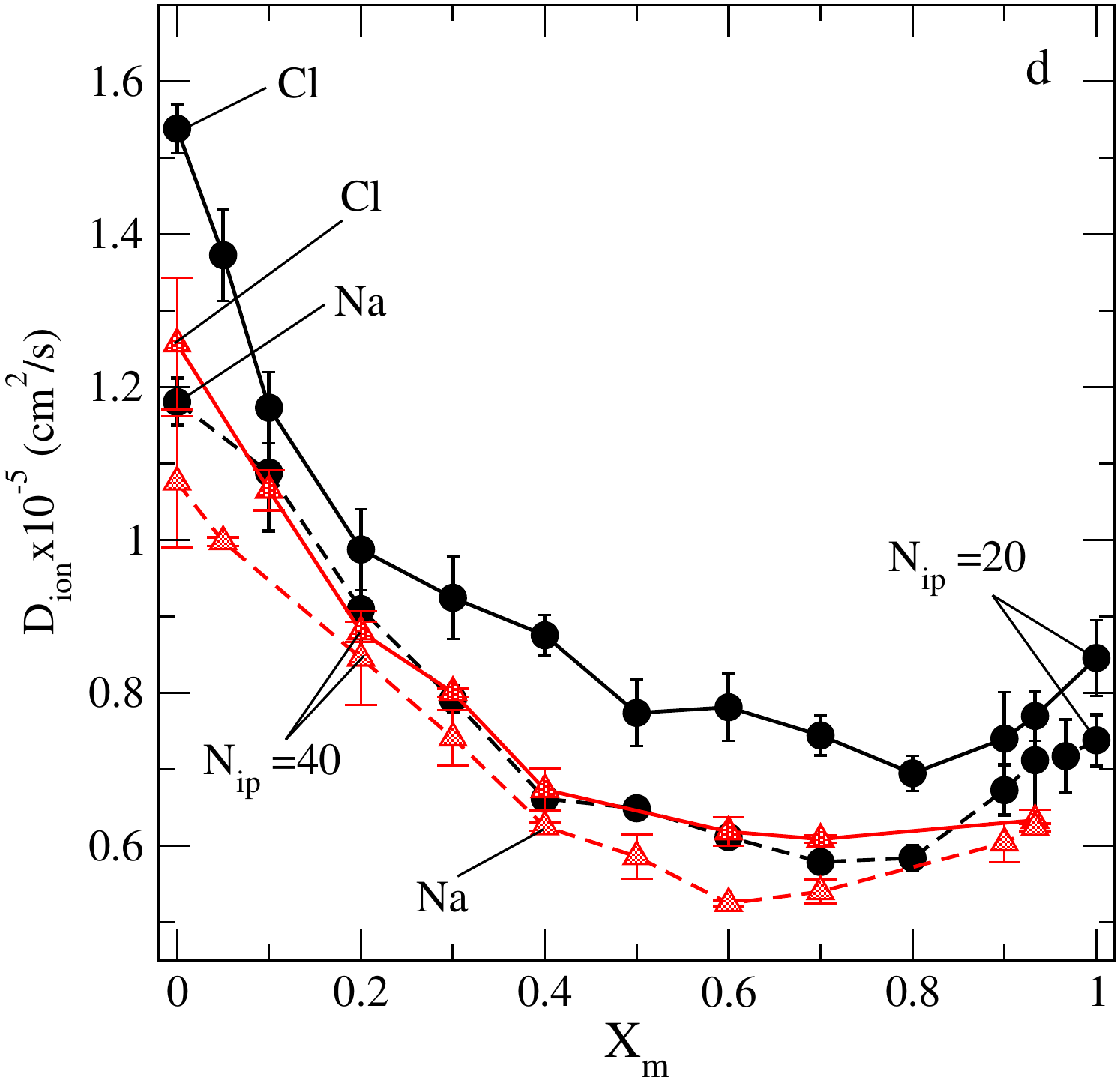}
\caption{\label{fig12}(Colour online) Changes of the self-diffusion coefficients of species
upon changing the ion fraction and solvent composition.}
\protect
\end{figure}

The self-diffusion coefficients of ions, i.e., water and methanol, in our work  were calculated 
 from the mean-square displacement (MSD) of a particle via Einstein relation,
\begin{equation}
D_i =\frac{1}{6} \lim_{t \rightarrow \infty} \frac{\rd}{\rd t} \langle\vert {\bf 
r}_i(\tau+t)-{\bf r}_i(\tau)\vert ^2\rangle,
\end{equation}
where $i$ refers to water or methanol, or ions Na, Cl,
and $\tau$ is the time origin. 
Default settings of GROMACS were used for the separation of the time origins. 
Moreover, the  fitting interval (from 10\% to 50\% of the 
analyzed trajectory) has been used to calculate $D_{\text m}$ and $D_{\text w}$.
Also, a special care has been taken regarding the fitting for the cases with a small 
number of particles on the extremes as well as of cations and anions
along the $X_{\text m}$ axis. 
A set of our results is given in  figure~\ref{fig12}. 

First, it is worth mentioning that we performed additional calculations for
the salt-free system in order to establish the magnitude of the effects due to the presence
of ionic solutes.
The $D_{\text w}$ is lower for all systems with ions ($X_{\text{ion}} \neq 0$) compared to 
the salt-free system ($X_{\text{ion}} = 0$) in the entire solvent composition interval,
upper left-hand panel in figure~\ref{fig12}. The result for $D_{\text w}$, in a salt-free system at $X_{\text m}=0$, is 
undoubtedly correct, which is around 2.6 (for SPC/E model).
A bigger amount of ion solutes leads to a reduction of
the value for $D_{\text w}$. However, a weaker effect of this sort is observed in water-rich
solvents, apparently  because there is enough water molecules to provide solvation of 
all the ions in the system and still there is enough ``free'' water. A more pronounced
reduction of $D_{\text w}$ is observed in the systems at intermediate values of $X_{\text m}$, because
practically all water molecules belong to the solvation shell of ions. 

As concerns the self-diffusion coefficient of methanol species (right-hand upper panel in figure~\ref{fig12}) 
we would like to mention the following trends. For the salt-free system, the value 
for $D_{\text m}$ is correct, which is around 2.1 at $X_{\text m}=1$. The ions are solvated by methanol molecules in the
methanol abundant region of the solvent compositions. 
Consequently, $D_{\text m}$ is lower for all such solutions in comparison with
the salt-free system. If the number of solutes is low ($N_{\text{ip}}=20$), the 
self-diffusion coefficient $D_{\text m}$ in the solution is practically equal to the values
obtained for the salt-free system in case $X_{\text m}$ is low (in the water-rich solutions). In other
words methanol molecules do not permeate the solvation shells of ions at these conditions.
Nevertheless, a reduction of the values for $D_{\text m}$ is pronounced if the amount of solute ions
increases from $N_{\text{ip}}=20$ to $N_{\text{ip}}=40$ and 60.  In particular, if the number of
methanol molecules is low, i.e., at low values of $X_{\text m}$, the reduction of $D_{\text m}$ is due to 
hydrogen bonds between methanols with water molecules in the solvation shells 
as well as due to methanols ``directly'' participating in the solvation of ions. 
It is impossible to precisely evaluate
the contributions from each of the two factors separately.  

An overall picture of the dependence of self-diffusion coefficients of all species in
the solution on the composition is shown in the lower left-hand panel in figure~\ref{fig12}. The behaviour of $D_{\text w}$ and $D_{\text m}$ 
on composition resembles the trends already observed for the salt-free mixtures with a minimum 
of each function in the region of intermediate $X_{\text m}$, see, e.g.~\cite{ewa3,galicia2}.
On the other hand, we are able to definitely state that $D_{\text{Cl}}$ and $D_{\text{Na}}$ are both much lower
than $D_{\text w}$ and $D_{\text m}$. Moreover, we are convinced that $D_{\text{Cl}}$ is higher than $D_{\text{Na}}$
in the entire interval of $X_{\text m}$, and, in particular, in the limit of pure water as a solvent.
The results obtained after the runs of 90~ns still do not provide quite accurate values.
In spite of the error margins, we note that both $D_{\text{Cl}}$ and $D_{\text{Na}}$ have a minimum
on a methanol-rich side of the solvent composition. To our best knowledge, 
an alternative method for 
calculation of $D_i$ via the velocity autocorrelation functions provides the
same accuracy.  It is surprising that the curves obtained by this method with only 
three ion pairs (and with 250 solvent molecules)
for the system in question, but with the use of a different model, are nearly 
perfect (see figure~8 of \cite{bouazizi}). Still another modelling employed by 
Hawlicka et al.~\cite{ewa3} with solely four data points for solutions with a mixed solvent
indicates that both $D_{\text{Cl}}$ and $D_{\text{Na}}$ have a minimum at $X_{\text m}=0.5$. 
Therefore, it is clear that further studies of the systems with
a larger number of particles  are indispensable.

As concerns the effect of ion concentration, we attempted to evaluate the self-diffusion 
coefficients with a larger number of ions, $N_{\text{ip}}=40$, and to make comparisons with the
previous case, $N_{\text{ip}}=20$. These results are shown in the right-hand lower panel in figure~\ref{fig12}.
Both, the $D_{\text{Cl}}$ and $D_{\text{Na}}$, decrease in magnitude at a higher ion fraction. The
reason of such a behaviour is in a higher density of well-solvated ions in the system.
Consequently, their movement is more hindered.   

The last part of this subsection is concerned with a few  peculiar issues of the 
dielectric response of NaCl solutions with water-methanol mixed solvents. 
Since an electrolyte solution is a conductor, the mobile ions move in response to 
an imposed static electric field. Thus, while dealing with this type of systems,
it is common to define a static equilibrium property that plays a role similar to
the dielectric constant of an insulating fluid, see, e.g., a comprehensive discussion of
the problem in \cite{hubbard}. Actually, the apparent dielectric constant reported from 
the experimental measurements  is an extrapolation from the frequency dependent
dielectric constant, as documented, for example, in \cite{kaatze} while discussing 
the dielectric properties of NaCl aqueous solutions of different salinity.
However, one should have in mind that there exist additional dynamic contributions
hidden in the extrapolated values. In several publications that describe the
application of molecular dynamics computer simulations to electrolyte solutions,
the static dielectric constant is defined or estimated in terms of the time-average of the 
fluctuations of the total dipole moment of the system \cite{caillol,payne,sala,zasetsky}, 
obtained by the summation of the dipole moment vectors of the entire set of water and 
methanol molecules in the simulation box (see, e.g., the GROMACS manual available online
and \cite{gromacs}), 
\begin{equation}
\varepsilon=1+\frac{4\piup}{3k_\text{B}TV}\big(\langle\bf M^2\rangle-\langle\bf M\rangle^2\big),
\label{eq3}
\end{equation}
where $k_\text{B}$ is Boltzmann's constant and $V$ is the simulation cell volume. 
The average value of all Cartesian projections of $\langle\bf M\rangle$ should be small
to provide the evidence that particles on the scale of the box 
are distributed homogeneously, see, for example,  figure~5~(c) of \cite{wasser}
for illustration of similar calculations for pure water.
We use the procedure based on equation~(\ref{eq3}) to get an insight into the
predictions of the model that involves a specific set of force fields (for ions, water
and methanol).

As discussed in detail in \cite{krienke} for aqueous electrolyte solutions with
completely dissociated solute molecules (NaCl in water is an example, see
table~7.6 of \cite{krienke}), the static permittivity of the sample, $\varepsilon$, 
decreases with increasing the electrolyte concentration. This permittivity simultaneously
represents the static permittivity of the solvent in the solution. Moreover,
the concentration dependent decrease of the static solvent permittivity is
defined as the dielectric depression and is empirically well fitted by
the linear dependence. In the present case, the situation is more complex since
the dielectric depression is determined by the solvent composition, apart from the
ion concentration. Furthermore, at the moment we are unaware how sensitive 
the dielectric depression is to the details of the force fields for water, methanol
and ions. 

\begin{figure}[!t]
\centering
\includegraphics[width=0.47\textwidth]{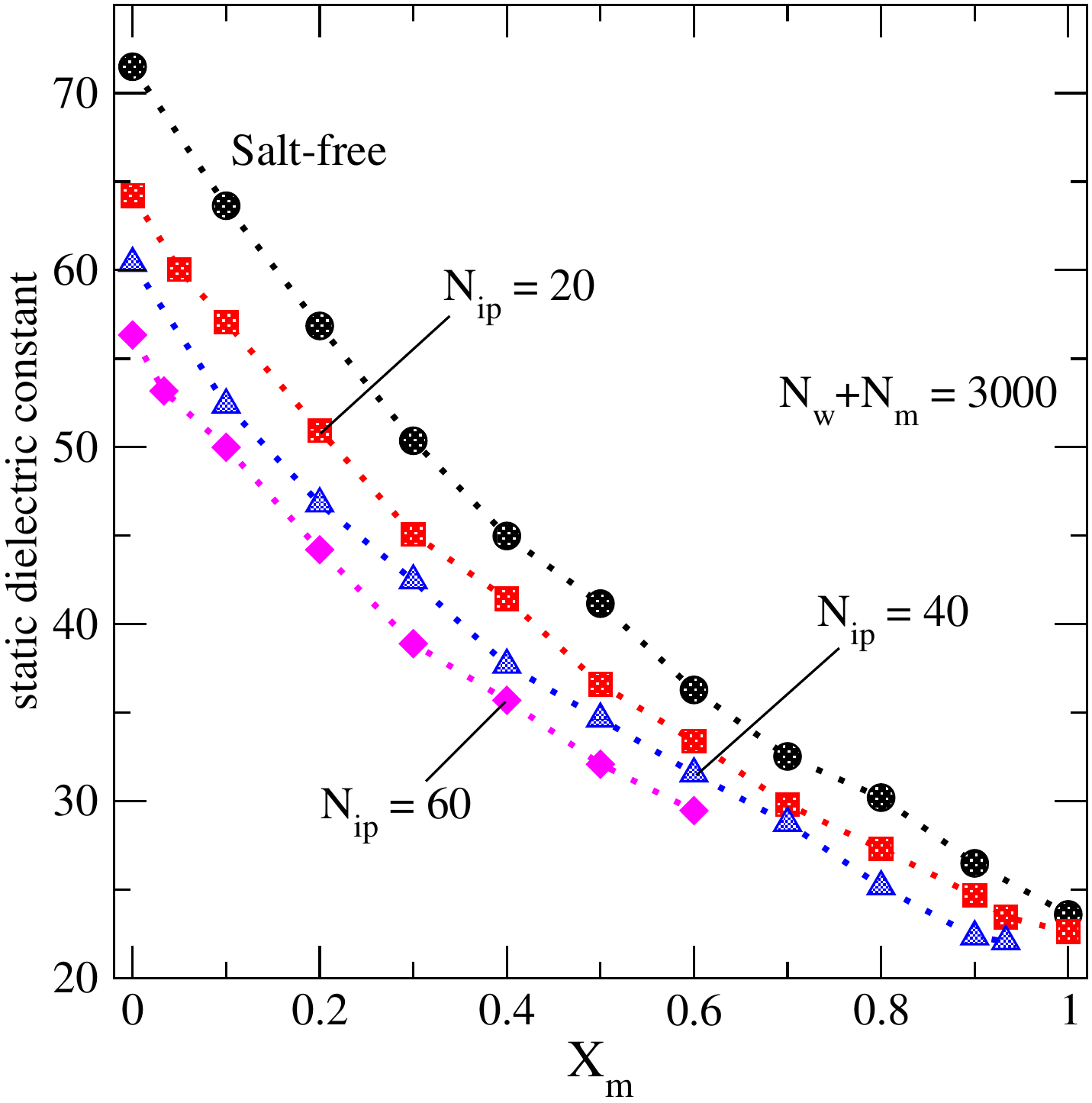}
\caption{\label{fig13}(Colour online) Composition dependence of the dielectric constant of ionic
solutions with water-methanol  mixed solvent at different ion fractions
from the simulations of the present work.}
\protect
\end{figure}

The curves from simulations for a combined water-methanol solvent and 
for the three solutions at different ion fractions are shown in figure~\ref{fig13}.
As it follows from the comparisons of the dielectric constant of a salt-free system  
and the experimental data performed in the recent work from this laboratory~\cite{galicia2},
the solvent model underestimates the values for $\varepsilon$ in the
entire interval of the composition. Nevertheless, the curve behaves qualitatively correctly, 
and the discrepancies between the simulation results  and experimental data are not big,
see figure~4 of \cite{galicia2}. 
Principal effects of ions can be seen on 
the water-rich side. Namely,
the dielectric constant decreases in magnitude with an increasing ion fraction. 
This behaviour is qualitatively
correct. On the other hand, on the methanol-rich side, all three curves describing 
the systems with different ion fractions flow closer, indicating a weaker response
of the systems with a predominant number of methanol molecules. 
Undoubtedly, it is of interest to extend the study of dielectric properties along
various lines briefly discussed above in a specifically focused future research.

\section{Summary and conclusions}

We have performed an extensive set of molecular dynamic simulations in 
the NPT ensemble to study 
properties of NaCl solutions with water-methanol solvent in the entire range of a
solvent composition. 
We explored systems at a fixed number of solvent molecules and three
values of the number of ion solutes. The salt-free system has been discussed as well.
Out of many possible versions of the model for a solution, only one
was studied. We consider this as a first step of  more systematic studies
of solutions involving ionic solutes of different complexity and 
a solvent mixture of water and methanol at different thermodynamic
states using nonpolarizable models.

We explored the evolution of a microscopic
structure of all the systems in terms of the pair distribution functions together with
the first coordination numbers of the species.  We
approached saturation at certain thermodynamic conditions. The number of clusters and maximal cluster size as
a function of time have been explored. 
Statistical features of the hydrogen bonds
in a water subsystem and of water-organic co-solvent bonds have been described. 
The self-diffusion coefficients of all the species and the 
dielectric constant were also calculated. All the simulations were performed at 
room temperature and at ambient pressure, 1~bar. 

From a comparison with a limited set of experimental data for the systems in question and
with the results of other authors on the related systems, we can conclude that the predictions
obtained are qualitatively correct and give a physically sound picture of the properties explored.
We observed that practically all the properties investigated are sensitive to the
composition of a water-organic liquid solvent. 

At the present stage of the development in the field, the missing and very important elements are really numerous
and need much more detailed investigations.
Namely, it is necessary to investigate  other force fields for methanol species at the united atom
and/or  all atom level of modelling for the type of solutions of this work. 
A comprehensive discussion of the modelling strategies that focused onto
a better understanding of aqueous solutions has been recently presented by Nezbeda et al.~\cite{ivo}.
The role of combination rules should be explored as well, see, e.g.,~\cite{spohr} for
the case of aqueous solutions of NaCl.
A possibility for fitting parameters of different models to avoid a spontaneous 
cluster formation in accordance with experimental findings for stable
solutions with composite a water-methanol solvent has not been explored so far,
in contrast to simpler aqueous solutions~\cite{aragones}.
It is extremely desirable to perform comparisons with structure factors coming from experiments 
(we are not aware of the experimental data for NaCL solutions with water-methanol solvents)
in close similarity to the developments available for salt-free mixtures, e.g.,~\cite{galicia1},
and, e.g., for aqueous solutions of salts with a different 
complexity~\cite{ariel,vega,bellissent,jungwirth}.
On the other hand, trends of the behaviour of thermodynamic, and, in particular, mixing properties
require further investigations. We hope to address some of these issues in our future studies.
The present work has partly paved the way toward the understanding of what should be done. 

\section*{Acknowledgements}
O.P. is grateful to D. Vazquez and M. Aguilar for technical support of this work
at the Institute of Chemistry of the UNAM.  M.C.S. is grateful to CONACyT for 
support of the Ph.D. studies.

\ukrainianpart

\title{Вплив концентрації іонів та складу розчинника на властивості водно-метанолових розчинів NaCl. 
Результати комп'ютерного моделювання методом молекулярної динаміки в NPT ансамблі} 

\author{M. Круз Санчес\refaddr{label1}, Дж. Гуйт\refaddr{label2}, С. Соколовскі\refaddr{label3}, O. Пізіо\refaddr{label4}}

\addresses{
\addr{label1} Інститут хімії, Національний автономний університет Мексики,
м. Мехіко, Мексика
\addr{label2} Відділ хімії, університет м. Падерборн, Німеччина
\addr{label3}Відділ моделювання фізико-хімічних процесів, університет Марії Кюрі-Склодовської,  Люблін, Польща
\addr{label4} Інститут матеріалознавства, Національний автономний університет Мексики,
м. Мехіко, Мексика 
}

\makeukrtitle

\begin{abstract}
Для дослідження  мікроскопічної структури та інших властивостей модельного розчину, що складається з солі  NaCl, розчиненої у суміші 
води та метанолу, використано ізотермічно-ізобаричне моделювання  методом молекулярної динаміки.
Щоб описати всю систему, модель води  SPC/E  і  модель еквівалентного атома  для метанолу поєдані  із силовим полем для іонів згідно з Дангом 
[\,J. Amer. Chem. Soc.,  1995, \textbf{117}, 6954].  
Нашою головною метою є дослідження впливу двох змінних, а саме, складу розчинника та концентрації іонів на густину розчину, на структурні 
властивості, на коефіцієнти самодифузії сортів та на діелектричну сталу. Крім того,
здійснено детальний аналіз перших координаційних чисел сортів. Визначено тенденції  поведінки  середньої кількості водневих зв'язків.

\keywords  водно-метанолові суміші, розчини електролітів, хлорид натрію, мікроскопічна структура, моделювання методом молекулярної динаміки

\end{abstract}

\end{document}